\documentclass[twocolumn]{aastex631}

\usepackage{hyperref}
\usepackage{soul}
\usepackage{multirow}
\usepackage[svgnames]{xcolor}

\hypersetup{linkcolor=CornflowerBlue,citecolor=CornflowerBlue,filecolor=CornflowerBlue,urlcolor=CornflowerBlue}
\defcitealias{t1e_1}{Allen \& Espinoza et al. 2024}
\defcitealias{t1e_2}{Allen \& Espinoza et al. 2025}
\shorttitle{JWST TRAPPIST-1~e/b Program}
\shortauthors{Allen et al.}
\begin{document}

\title{JWST TRAPPIST-1~e/b Program: Motivation and first observations}

\author[0000-0002-0832-710X]{Natalie H. Allen}\altaffiliation{NSF Graduate Research Fellow}
\affiliation{Department of Physics and Astronomy, Johns Hopkins University, 3400 N. Charles Street, Baltimore, MD 21218, USA}
\correspondingauthor{Natalie H. Allen}
\email{nallen19@jhu.edu}

\author[0000-0001-9513-1449]{N\'estor Espinoza}
\affiliation{Space Telescope Science Institute, 3700 San Martin Drive, Baltimore, MD 21218, USA}
\affiliation{Department of Physics and Astronomy, Johns Hopkins University, 3400 N. Charles Street, Baltimore, MD 21218, USA}

% data + theory help so far
\author[0000-0002-4945-1860]{V. A. Boehm}
\affiliation{Department of Astronomy and Carl Sagan Institute, Cornell University, 122 Sciences Drive, Ithaca, NY 14853, USA}

\author[0000-0003-4835-0619]{Caleb I. Ca\~nas}
\affiliation{Exoplanets and Stellar Astrophysics Laboratory (Code 667), NASA Goddard Space Flight Center, Greenbelt, MD 20771, USA}
\affiliation{Center for Research and Exploration in Space Science and Technology II, NASA/GSFC, Greenbelt, MD 20771, USA}

\author[0000-0002-7352-7941]{Kevin B. Stevenson}
\affiliation{JHU Applied Physics Laboratory, 11100 Johns Hopkins Rd, Laurel, MD 20723, USA}

\author[0000-0002-8507-1304]{Nikole K. Lewis}
\affiliation{Department of Astronomy and Carl Sagan Institute, Cornell University, 122 Sciences Drive, Ithaca, NY 14853, USA}

\author[0000-0003-4816-3469]{Ryan J. MacDonald}
\affiliation{School of Physics and Astronomy, University of St Andrews, North Haugh, St Andrews, KY16 9SS, UK}

\author[0000-0003-2528-3409]{Brett M.~Morris}
\affiliation{Space Telescope Science Institute, 3700 San Martin Drive, Baltimore, MD 21218, USA}

% further team contributions

\author[0000-0002-0802-9145]{Eric Agol}
\affiliation{Department of Astronomy and Virtual Planetary Laboratory, University of Washington, Box 351580, Seattle, WA 98195, USA}

\author[0000-0001-8020-7121]{Knicole Col\'{o}n}
\affiliation{NASA Goddard Space Flight Center, Greenbelt, MD 20771, USA}

\author[0000-0001-8274-6639]{Hannah Diamond-Lowe}
\affiliation{Space Telescope Science Institute, 3700 San Martin Drive, Baltimore, MD 21218, USA}

\author[0000-0002-5322-2315]{Ana Glidden}
\affiliation{Department of Earth, Atmospheric and Planetary Sciences, Massachusetts Institute of Technology, Cambridge, MA 02139, USA}
\affiliation{Kavli Institute for Astrophysics and Space Research, Massachusetts Institute of Technology, Cambridge, MA 02139, USA} 

\author[0000-0003-0854-3002]{Am\'{e}lie Gressier}
\affiliation{Space Telescope Science Institute, 3700 San Martin Drive, Baltimore, MD 21218, USA}

\author[0000-0001-5732-8531]{Jingcheng Huang}
\affiliation{Department of Earth, Atmospheric and Planetary Sciences, Massachusetts Institute of Technology, Cambridge, MA 02139, USA}

\author[0000-0003-0525-9647]{Zifan Lin}
\affiliation{Department of Physics and McDonnell Center for the Space Sciences, Washington University, St. Louis, MO 63130, USA}

\author[0000-0002-2508-9211]{Douglas Long}
\affiliation{Space Telescope Science Institute, 3700 San Martin Dr, Baltimore, MD 21218, USA}

\author[0000-0002-2457-272X]{Dana R. Louie}
\affiliation{Catholic University of America, Department of Physics, Washington, DC, 20064, USA}
\affiliation{Exoplanets and Stellar Astrophysics Laboratory (Code 667), NASA Goddard Space Flight Center, Greenbelt, MD 20771, USA}
\affiliation{Center for Research and Exploration in Space Science and Technology II, NASA/GSFC, Greenbelt, MD 20771, USA}

\author[0000-0001-7891-8143]{Meredith A. MacGregor}
\affil{Department of Physics and Astronomy, Johns Hopkins University, 3400 N. Charles Street, Baltimore, MD 21218, USA}

\author[0000-0003-3818-408X]{Laurent Pueyo}
\affiliation{Space Telescope Science Institute, 3700 San Martin Drive, Baltimore, MD 21218, USA}

\author[0000-0002-3627-1676]{Benjamin V.\ Rackham}
\affiliation{Department of Earth, Atmospheric and Planetary Sciences, Massachusetts Institute of Technology, 77 Massachusetts Avenue, Cambridge, MA 02139, USA}
\affiliation{Kavli Institute for Astrophysics and Space Research, Massachusetts Institute of Technology, Cambridge, MA 02139, USA}

\author[0000-0002-5147-9053]{Sukrit Ranjan}
\altaffiliation{Blue Marble Space Institute of Science}
\affiliation{Lunar \& Planetary Laboratory, University of Arizona, Tucson, AZ 85719}

\author[0000-0002-6892-6948]{Sara Seager}
\affiliation{Department of Physics and Kavli Institute for Astrophysics and Space Research, Massachusetts Institute of Technology, Cambridge, MA 02139, USA}
\affiliation{Department of Earth, Atmospheric and Planetary Sciences, Massachusetts Institute of Technology, Cambridge, MA 02139, USA}
\affiliation{Department of Aeronautics and Astronautics, MIT, 77 Massachusetts Avenue, Cambridge, MA 02139, USA}

\author[0000-0001-5455-6678]{Guadalupe Tovar Mendoza}
\altaffiliation{NSF MPS-Ascend Postdoctoral Fellow}
\affil{Department of Physics and Astronomy, Johns Hopkins University, 3400 N. Charles Street, Baltimore, MD 21218, USA}

\author[0000-0003-3305-6281]{Jeff A. Valenti}
\affiliation{Space Telescope Science Institute, 3700 San Martin Drive, Baltimore, MD 21218, USA}

\author[0000-0002-2643-6836]{Daniel Valentine}
\affiliation{University of Bristol, HH Wills Physics Laboratory, Tyndall Avenue, Bristol, UK}

\author[0000-0001-7827-7825]{Roeland P.~van der Marel}
\affiliation{Space Telescope Science Institute, 3700 San Martin Drive, Baltimore, MD 21218, USA}
\affiliation{Department of Physics and Astronomy, Johns Hopkins University, 3400 N. Charles Street, Baltimore, MD 21218, USA}

\author[0000-0003-4328-3867]{Hannah R. Wakeford}
\affiliation{University of Bristol, HH Wills Physics Laboratory, Tyndall Avenue, Bristol, UK}

%% Note that the \and command from previous versions of AASTeX is now
%% depreciated in this version as it is no longer necessary. AASTeX 
%% automatically takes care of all commas and "and"s between authors names.

%% AASTeX 6.31 has the new \collaboration and \nocollaboration commands to
%% provide the collaboration status of a group of authors. These commands 
%% can be used either before or after the list of corresponding authors. The
%% argument for \collaboration is the collaboration identifier. Authors are
%% encouraged to surround collaboration identifiers with ()s. The 
%% \nocollaboration command takes no argument and exists to indicate that
%% the nearby authors are not part of surrounding collaborations.

%% Mark off the abstract in the ``abstract'' environment. 
\begin{abstract}
One of the forefront goals in the field of exoplanets is the detection of an atmosphere on a temperate terrestrial exoplanet, and among the best suited systems to do so is TRAPPIST-1. However, JWST transit observations of the TRAPPIST-1 planets show significant contamination from stellar surface features that we are unable to confidently model. Here, we present the motivation and first observations of our JWST multi-cycle program of TRAPPIST-1~e, which utilize close transits of the airless TRAPPIST-1~b to model-independently correct for stellar contamination, with the goal of determining whether TRAPPIST-1~e has an Earth-like mean molecular weight atmosphere containing CO$_2$. We present our simulations, which show that with the 15 close transit observations, we will be able to detect this atmosphere on TRAPPIST-1~e at $\Delta\ln\,Z=5$ or greater confidence assuming we are able to correct for stellar contamination using the close transit observations. We also show the first three observations of our program. We find that our ability to correct for stellar contamination can be inhibited when strong stellar flares are present, as flares can break the assumption that the star does not change meaningfully between planetary transits. The cleanest observation demonstrates the removal of stellar contamination contribution through an increased preference for a flat line over the original TRAPPIST-1~e spectrum, but highlights how minor data analysis assumptions can propagate significantly when searching for small atmospheric signals. This is amplified when using the signals from multiple planets, which is important to consider as we continue our atmospheric search.

\end{abstract}

%% Keywords should appear after the \end{abstract} command. 
%% The AAS Journals now uses Unified Astronomy Thesaurus concepts:
%% https://astrothesaurus.org
%% You will be asked to selected these concepts during the submission process
%% but this old "keyword" functionality is maintained in case authors want
%% to include these concepts in their preprints.

%% From the front matter, we move on to the body of the paper.
%% Sections are demarcated by \section and \subsection, respectively.
%% Observe the use of the LaTeX \label
%% command after the \subsection to give a symbolic KEY to the
%% subsection for cross-referencing in a \ref command.
%% You can use LaTeX's \ref and \label commands to keep track of
%% cross-references to sections, equations, tables, and figures.
%% That way, if you change the order of any elements, LaTeX will
%% automatically renumber them.
%%
%% We recommend that authors also use the natbib \citep
%% and \citet commands to identify citations.  The citations are
%% tied to the reference list via symbolic KEYs. The KEY corresponds
%% to the KEY in the \bibitem in the reference list below. 

\section{Introduction} \label{sec:intro}
%Despite incredible progress in the detection and characterization of exoplanet atmospheres over the past two decades, we have not yet definitively detected the atmosphere of a terrestrial exoplanet. This is not for lack of effort: the atmospheres of tens of terrestrial exoplanets have already been observed with JWST using transmission spectroscopy in its first few years of operation. Many of the resulting transmission spectra have been flat, making it difficult to say if the underlying planetary signal is due to a complete lack of atmosphere, or just a heavy atmosphere or high cloud deck making the features undetectable \citep[see e.g.][]{lustig-yaeger_2023, may_2023, kirk_2024}. Stellar contamination, or signals injected into transmission spectra due to the presence of active regions on the host star's surface, has also been a complicating factor -- when we observe features in the spectrum, it can be difficult to tell if their origin is planetary or stellar \citep{moran_2023, lim_2023, Radica_2024}. There have been a few (I think this is boring actually)

Of the temperate terrestrial exoplanetary systems currently feasible for atmospheric detection, the TRAPPIST-1 system stands out as one of the most promising. Not only does it have seven terrestrial planets orbiting an M dwarf star only about as large as Jupiter \citep[$R_s = 0.1192\pm0.0013\,R_\odot$,][]{Agol_2021}, but it has multiple planets that lie within the ``habitable zone" of the star. The community has made sure to take advantage of this opportunity: over 400 hours of JWST observations have already been devoted to observe all seven planets in the system \citep{Espinoza_2025}. Early observations in emission with MIRI show that the innermost planet, TRAPPIST-1~b, is unlikely to host a significant atmosphere, with results for TRAPPIST-1~c more uncertain \citep{greene_2023, Ih_2023, zieba_2023, Ducrot_2025, Gillon_2025, Maurel_2025}. %The planets beyond TRAPPIST-1~c become too cool to be easily observable in emission (this happens quite abruptly due to the T$^4$ relation of emissive flux), and therefore have been mostly observed in transit instead. 
Observations in transmission, however, have been more difficult to interpret. M dwarfs are known to be generally magnetically active, with plenty of evidence of rotational modulation due to stellar surface active regions rotating in and out of view \citep[e.g.,][]{Newton_2017}. Even among the M dwarf class, TRAPPIST-1 is a particularly active star, with evidence of both hot and cold surface features that evolve on timescales of about a stellar rotation period ($P_{rot} = 3.3$ days) timescales \citep{Morris_2018}, as well as regular flare events \citep[with predictions for flares of energy $10^{30}$ erg in the TESS bandpass, large enough to impact transit spectra, to occur multiple times a day,][]{Howard_2023}. The presence of active regions on the stellar photosphere cause an effect termed the ``transit light source effect," or stellar contamination more generally \citep[see e.g.][]{Sing_2011,Rackham_2018, Rackham:2019}. This can inject slopes and even molecular features into the transmission spectrum that can look like potential features from the planet even though they originate from the star, due to the difference in light source in and out of the transit chord. This effect has been inescapable in JWST observations of the TRAPPIST-1 system, where observations of the likely airless TRAPPIST-1~b, as well as planets c and d, show stellar contamination signals much larger than any predicted atmospheric signal size \citep{lim_2023, Radica_2024, Piaulet-Ghorayeb_2025}.

The problem of stellar contamination persists far beyond the TRAPPIST-1 system and has been a significant complicating factor in the search for an atmosphere on a rocky exoplanet, for which we currently have no conclusive evidence \citep[see][for a recent review of efforts by JWST for characterizing exoplanet atmospheres] {Espinoza_2025}, though there has been recent evidence for some ultra hot (T$_{eq}>2000$ K) planets with volatile atmospheres maintained through their molten surfaces \citep[55 Cancri~e and TOI-561~b,][]{Hu_2024, Patel_2024, Teske_2025}. However, this complication does not overshadow the significant observational chance provided by the ``M dwarf opportunity" \citep[e.g.,][]{Charbonneau_2007}, nor the importance of understanding the nature of the most common stars, and thus rocky planet hosts, in the universe \citep[see e.g., discussion and reasoning in][]{Redfield_2024}. Therefore, our studies of rocky planets orbiting M dwarfs will persist despite the complication presented by the stars.

In order to work towards an answer to the question of the state of M dwarf rocky planets, we must tackle stellar contamination. Unfortunately, the existing methods to correct stellar contamination are insufficient to robustly recover the planet's atmosphere. The typical framework is to assume that some fraction of the host star's photosphere is replaced by a surface of another photospheric temperature corresponding to the spot temperature \citep{Rackham_2018, Pinhas_2018, Espinoza_2019, Rackham_2023}. It is well known that this assumption is an oversimplification \citep[see e.g.,][]{Witzke_2022,Norris_2023, Smitha_2025}. Magnetic and three-dimensional effects have been suggested to be important and have a significant impact on the resulting spectrum, which has already been seen in early magnetohydrodynamic surface feature models for stars of multiple spectral types \citep[e.g.][]{Witzke_2022, Smitha_2025}. For M dwarfs, our ability to model and understand all components contributing to the stellar spectra we observe is especially lacking, and has already been seen for the TRAPPIST-1 system in particular \citep[][]{lim_2023, Radica_2024, Espinoza:2025}. Accurate surface feature models are underway, but require years of development and benchmarking before they are feasible for use. 

In the meantime, there are efforts towards correcting for stellar contamination without surface feature models. One of the most promising is the use of close transits of multiple planets \citep{TJCI_2024, Rathcke_2025}. Since stellar contamination is a multiplicative effect, the contamination signals should be the same for planets illuminated by the same stellar surface as long as the contamination is coming from unocculted spots or the two planets have the same transit chord. Therefore, if multiple planets in the same system are observed in close succession, such that we can assume the stellar surface is constant between the two transits (though in reality this scenario may be complicated by effects such as rotation of surface features on and off the surface and feature evolution in time), the two transmission spectra can be used together to correct for the stellar contamination in a model-independent fashion. Ideally, if one of the planets has no atmosphere and thus no features in transmission, the stellar contamination does not change between transits, and the two planets have the same occulted and unocculted contamination, the ratio of the two spectra will result in the clean spectrum of the planet with an atmosphere, simply normalized by the constant transit depth of the airless planet. 

The TRAPPIST-1 system is an ideal test for the use of close transits to correct for stellar contamination \citep[e.g.,][]{Rathcke_2025}, since the innermost planet in the system, TRAPPIST-1~b, likely has no atmosphere, and the system hosts multiple planets in the habitable zone. Earlier observations of TRAPPIST-1~e as part of the TST-DREAMS GTO program \citep{Espinoza:2025, Glidden_2025} show signs of features, some of which may be attributable to a planetary atmosphere, but which are dwarfed in size by stellar contamination that cannot be corrected for using current stellar models. Motivated by this, we proposed to utilize the method of close transits for the model-independent removal of stellar contamination by using TRAPPIST-1~b as a stellar contamination proxy for TRAPPIST-1~e, and we were awarded approximately 130 hours in JWST Cycles 3 and 4 to observe 15 close transits in order to determine whether TRAPPIST-1~e hosts an Earth-like atmosphere (or an atmosphere of similar molecular weight) \citepalias[GO 6456 and 9256,][]{t1e_1, t1e_2}.

This paper describes the motivation for our JWST program and showcases the first observations. We provide the simulations that motivated the program and the program setup in \autoref{sec:motivation}. In \autoref{sec:obs}, we present the first three observations, taken in 2024. We discuss the results of these observations in \autoref{sec:discussion} before finishing with our conclusions in \autoref{sec:conclusion}.

\section{Program Motivation}\label{sec:motivation}
%As previously stated, the TRAPPIST-1 system is perfect for the testing of the use of close transits to model-independently correct for stellar contamination, as it has been shown that the internal planet TRAPPIST-1~b is likely to have no atmosphere \citep{greene_2023, Ducrot_2025}. On the other hand, t
While TRAPPIST-1~b is likely to be airless \citep{greene_2023, Ih_2023, Ducrot_2025, Gillon_2025, Maurel_2025}, the atmospheric states of the TRAPPIST-1 planets on longer orbital periods is currently unknown, beyond constraints from (a priori unlikely) scenarios of H/He dominated light atmospheres that can be ruled out from past observations \citep[e.g.,][]{deWit_2016, Zhang_2018, deWit_2018, Wakeford_2019, Garcia_2022, Espinoza:2025, Berardo_2025}. Of particular interest are the planets that lie in the habitable zone, as TRAPPIST-1 provides arguably the only opportunity with current observational capabilities to observe the atmosphere of a true temperate, terrestrial habitable zone planet. TRAPPIST-1~e is especially interesting as a test of atmospheric evolution, survival, and habitability for temperate rocky planets around M stars due to its similarity in size and temperature to our own solar system terrestrial planets \citep[e.g.,][]{Gillon_2017}. Therefore, for our observations, we test the use of close transits of TRAPPIST-1~b and TRAPPIST-1~e to detect an Earth-like atmosphere on TRAPPIST-1~e. 

\subsection{Simulations}\label{sec:sims}
We carried out simulations to determine how many transit observations are needed to detect an Earth-like atmosphere on TRAPPIST-1~e. We create the spectrum of an Earth-like atmosphere with 1 bar surface pressure containing N$_2$ (background gas), H$_2$O, CO$_2$, CH$_4$, O$_2$, O$_3$, N$_2$O, and CO using POSEIDON \citep{poseidon1, poseidon2}, based on the composition presented in \citet{Lin_2021} for a modern Earth-like scenario on TRAPPIST-1~e. Note that this model contains 100x the CO$_2$ present on Earth today to keep the surface above freezing, but this does not meaningfully affect the detectability of the CO$_2$ feature as the feature strength is saturated at Earth concentrations. Using this as our base model, we simulate realistic observations of TRAPPIST-1~e and b.   

We ran stellar contamination retrievals of the level of active regions present on TRAPPIST-1 during observations of TRAPPIST-1~b \citep{lim_2023}, for which one observation is explainable by a surface dominated by hot spots and the other by cold spots. These are similar to the active regions found in retrievals of observations of TRAPPIST-1~e \citep{Espinoza:2025} and therefore seem to be somewhat indicative of the typical level and characteristics of activity on TRAPPIST-1 (approximately 2000 K and 2900 K for cold and hot spots, respectively, represented by PHOENIX models). Note that while we believe the PHOENIX models are not fully representative of true surface features, more complete models are not available across the wide range of temperatures we need to represent a variety of different surface features currently. In addition, as our simulations assume identical contamination signatures for both planets, such that the exact appearance of the contamination is less important than in reality.

With these hot and cold spot models, we create many instances of TRAPPIST-1 by assigning a fractional value of surface coverage outside of the transit chord taken up by hot and cold spots using a uniform prior from 0 to 1 using the formulation from \citet{Rackham_2018}. For each of these TRAPPIST-1 instances, we then simulate transmission spectra of TRAPPIST-1~b with no atmosphere, and TRAPPIST-1~e with the described Earth-like atmosphere, using empirical error bars on the transmission spectrum and the same wavelength binning scheme as previous observations of TRAPPIST-1~e with NIRSpec/PRISM from GTO 1331 \citep{Espinoza:2025}. We then repeat this process a number of times (10 for our retrieval test in \autoref{sec:complex}, 1000 for the simple test in \autoref{sec:simple}) to create multiple noise instances, and compute the depth ratio $R \equiv D_e/D_b$, where $D_b$ and $D_e$ are the transit depths of planets b and e, respectively.  

The error on the resulting transit depth ratio, $\sigma_R$, can be obtained through the delta method:
\begin{equation}\label{eq:err}
\frac{\sigma_R}{R} = \sqrt{\frac{\sigma_e^2}{D_e^2}+\frac{\sigma_b^2}{D_b^2}}
%    \sigma_R = \sqrt{\frac{\sigma_e^2}{D_b^2} + \frac{D_e^2}{D_b^4}\sigma_b^2}
\end{equation}
where $\sigma_b$ and $\sigma_e$ are the transit depth errors of planets b and e, respectively. One important corollary of this expression for the error $\sigma_R$ is that the signal-to-noise ratio of a planetary atmospheric feature $\Delta D_e$ in the transmission spectrum of planet e will always be larger than the corresponding signal-to-noise ratio of the same feature in the spectrum ratio $\Delta R = \Delta D_e / D_b$. With these simulated observations of the spectrum ratio, we perform a retrieval test to determine the number of observations necessary to find an Earth-like atmosphere on TRAPPIST-1~e (for a simpler test focused on our ability to detect the 4.3 $\mu$m CO$_2$ feature, with which we can perform for many more simulated noise instances due to the smaller computational cost, see \autoref{sec:simple}). 

\subsubsection{Atmospheric Retrievals with POSEIDON}\label{sec:complex}
Under the assumption that we are able to confidently remove the stellar contamination fully from our observations through the transmission spectra ratio method, there are many more potential spectroscopic features than CO$_2$ in the NIRSpec/PRISM bandpass. Therefore, we test our ability to identify an Earth-like atmosphere through full atmospheric retrievals. 

We modify \textit{POSEIDON} to run retrievals on the spectrum ratio. For the airless TRAPPIST-1~b, we assume a constant fixed transit depth of 7100 ppm, and therefore retrieve only the atmospheric composition of TRAPPIST-1~e. The free parameters in the retrieval are the abundances of each atmospheric species listed in \autoref{sec:sims} with a large log uniform prior range [-12,0], besides the background gas N$_2$ which is automatically taken as the remainder of the atmospheric composition such that, together with the other species, the sum of abundances is unity. Note that we also tested the use of centered-log-ratio priors for our atmospheric species \citep{Benneke_2012}, but did not see a significant difference in results. The resulting posteriors are more difficult to interpret due to the shape of the prior, though, so we focus on the results from the log uniform case. We also retrieve the atmospheric temperature (uniform between 50 and 450 K) and reference radius (uniform between 0.85$R_p$ and 1.15$R_p$, with $R_p=0.92R_\oplus$), for a total of nine free parameters. 

\begin{figure}
    \centering
    \includegraphics[width=\linewidth]{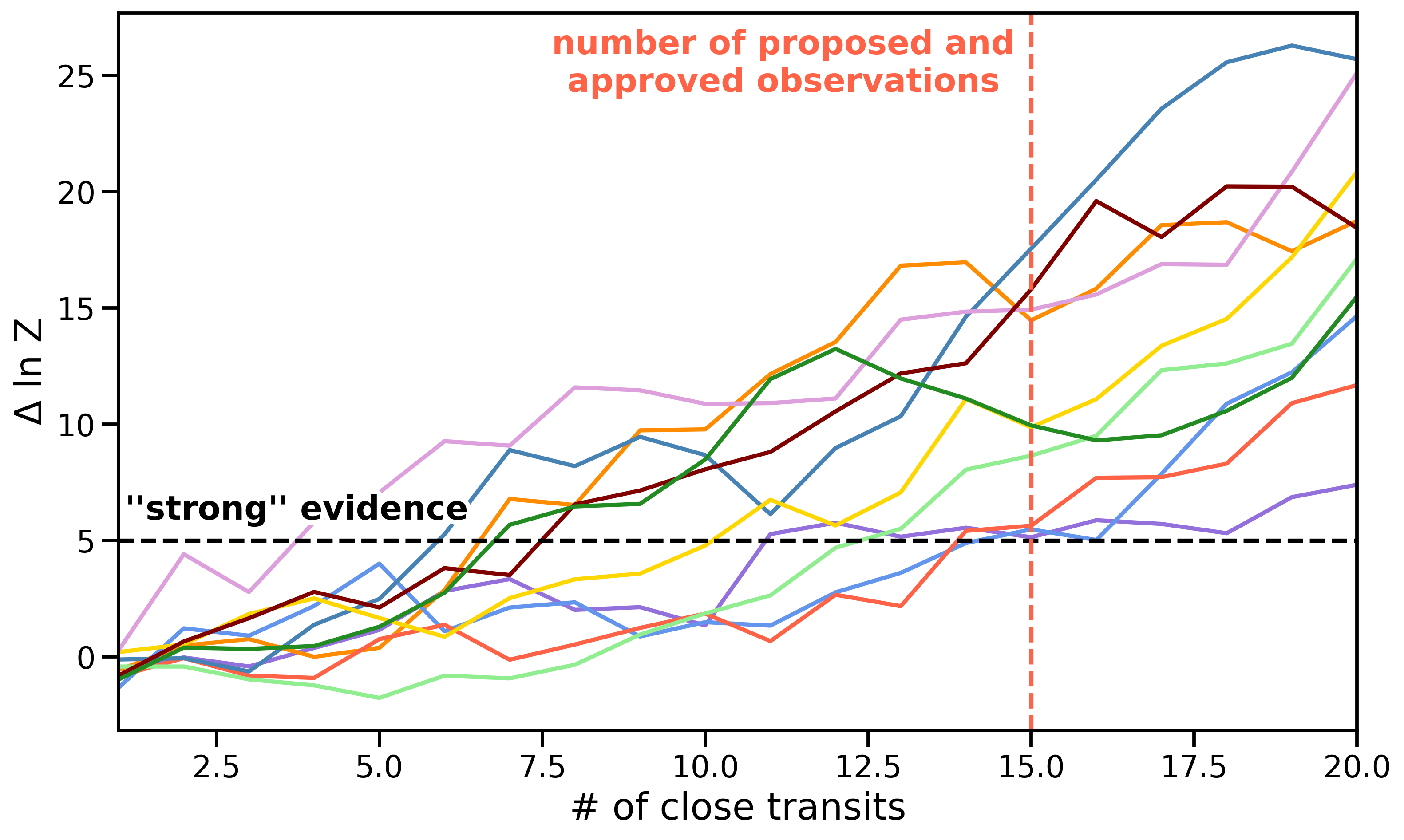}
    \caption{The atmospheric detection significance as a function of the number of observed close transits for a full Earth-like atmosphere retrieval with \textit{POSEIDON}. The 10 different noise instances are each given by a different color. With 15 close transits, we are able to achieve at least a $\Delta \,\,ln\,\,Z = 5$, ``strong evidence" detection of an Earth-like atmosphere. }
    \label{fig:complex-detect}
\end{figure}

We carried out these retrievals on ten different stellar contamination noise instances as described in \autoref{sec:sims}, considering one to 20 sets of transits. For each of these simulated observations, we also retrieve a flat line model, which is equivalent to our inability to detect an atmosphere. The difference in $\ln\,\,Z$, where Z is the Bayesian evidence, between these models determines our atmospheric detection capability. We follow the Jeffreys' scale for confidence: A $\Delta \,\,ln\,\,Z$ of greater than 2.5 is ``moderate", and of greater than 5 is ``strong" evidence \citep[at best,][]{Trotta:2008}. \autoref{fig:complex-detect} shows the $\Delta\,\, ln\,\,Z$ as a function of transit number for each of our ten noise instances.  

All of our noise instances are able to detect an atmosphere on TRAPPIST-1~e with 15 close transits with at least $\Delta \,\,ln\,\,Z=5$, or ``strong", significance, and often higher. However, beyond this agreement, which also matches our finding from \autoref{sec:simple}, our noise instances show a variety of outcomes. Some noise instances are able to ``find" an atmosphere (and at relatively high significance) earlier. Some of this is inherent to the fact that these retrievals consider the information content of multiple spectroscopic bands rather than just the 4.3 $\mu$m CO$_2$ feature as in our previous experiment, which boosts the detection significance for all of our retrievals. Regardless of the specific observational instances of noise, we are able to detect an atmosphere at $\Delta \,\,ln\,\,Z=5$ in any case presented with our 15 transit observations. %However, much of the additional difference in detection significance between realizations is just due to the inherent noise in our observations. When stacking this many observations together, we are more sensitive to the individual noise instances, each of which has an uncertainty larger than the atmospheric signals we are attempting to detect. This, in combination with the fact that relatively small outliers can sometimes drive atmospheric detections in retrievals due to their sensitivity in specific wavelength regions, we believe is the reason for the spread in detection sensitivities. The point remains, however, that 

\begin{figure}
    \centering
    \includegraphics[width=\linewidth]{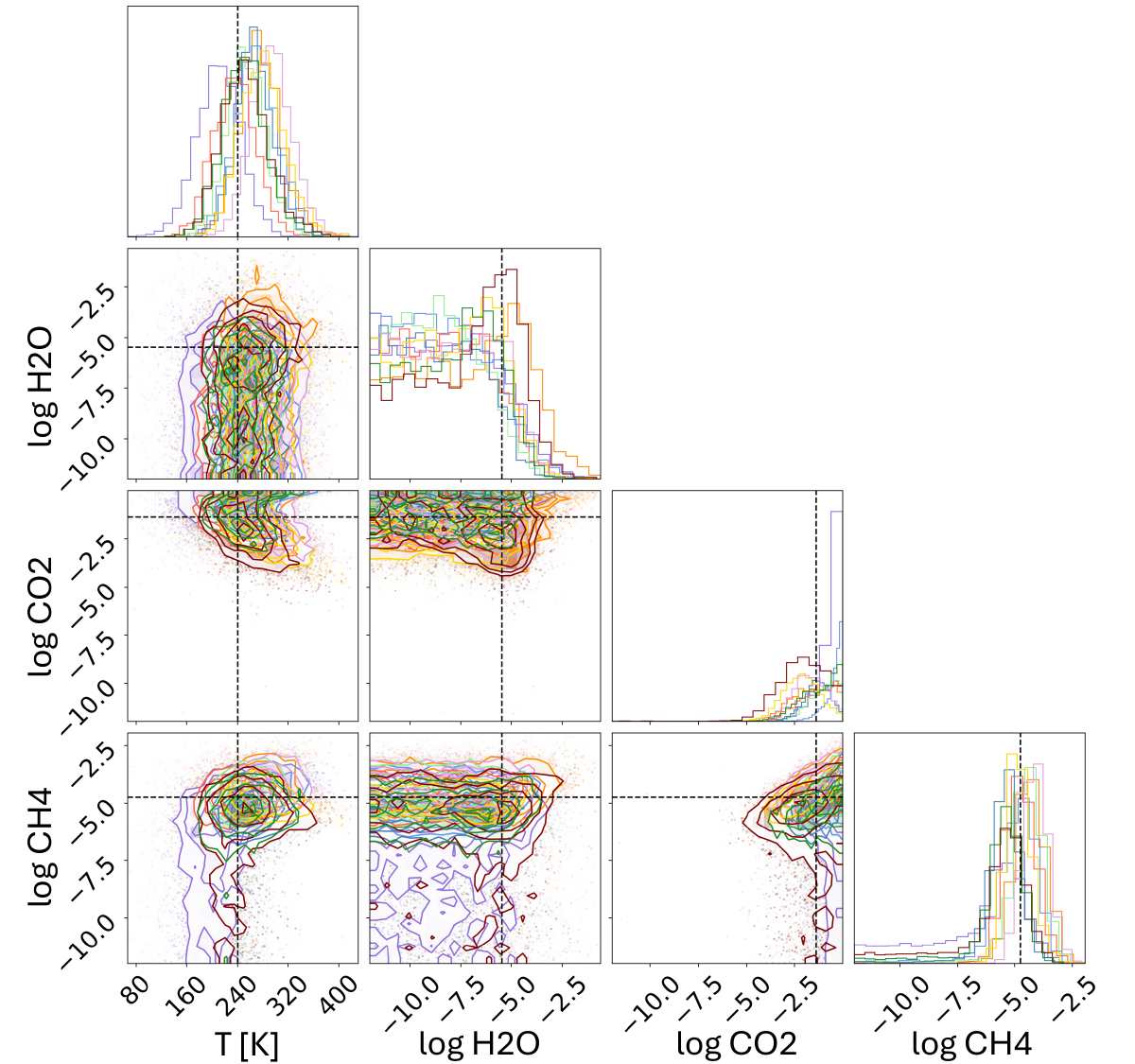}
    \caption{The posterior distributions associated with the 15 close transit retrievals for each of the 10 noise instances shown in \autoref{fig:complex-detect}. %The narrow posteriors in black correspond to zero noise retrievals, essentially the limit of detection significance given observations at the resolution of NIRSpec/PRISM and therefore taken as the ``ground truth" values. 
    We confidently detect CO$_2$ in all noise instances, and sometimes are able to detect CH$_4$ as well, while we do not detect (and therefore do not rely on detecting) H$_2$O.}
    \label{fig:corner}
\end{figure}

\autoref{fig:corner} shows the corresponding posteriors associated with the 15 close transit retrievals for the parameters of main interest, which are T and the abundances of H$_2$O, CO$_2$, and CH$_4$. The full corner plot is shown in \autoref{app:corner}. Each noise instance is given the same color as the evidence curve in \autoref{fig:complex-detect}. %In addition to the ten noise instances, shown with the same color as in \autoref{fig:complex-detect}, we also carry out a retrieval with data points with no scatter and no associated error bars to test the inherent information content of the spectrum -- i.e. to see the maximum possible constraints that the retrieval is able to obtain from a given spectrum at the resolution of NIRSpec/PRISM. This is taken to be our ``ground truth" (which has excellent agreement with the input values), and are shown in black in \autoref{fig:corner}.  
We confidently detect CO$_2$ in all of the noise instances. In addition, we are sometimes able to detect CH$_4$. We are unable to confidently detect all other species, which is overall unsurprising given the lack of spectroscopic features that many of these species have in this wavelength range. It is somewhat surprising that we do not confidently detect H$_2$O, although it is encouraging that we do not rely on this feature for detection, since it is likely present in the spots on TRAPPIST-1 and therefore difficult to separate from potential stellar features (see e.g., stellar contamination retrievals for observation 1 in \autoref{app:ret_fits}). %We show an example noise instance and the corresponding retrieved spectrum in \autoref{fig:sim_spectrum}. 

%\begin{figure}
%    \centering
%    \includegraphics[width=\linewidth]{spectrum_15_noise4.png}
%    \caption{The retrieved spectrum associated with one of the noise instances for 15 close transits. All of the noise instances result in essentially identical models, and thus only one is shown. }
%    \label{fig:sim_spectrum}
%\end{figure}

\subsection{Observing program setup}
With 15 close transits, assuming the transfer of stellar contamination information from TRAPPIST-1~b to TRAPPIST-1~e works as theorized, we should be able to determine if TRAPPIST-1~e has an Earth-like atmosphere or not according to our tests. We used this as the basis to design our observing program. We note that this number of transits is a modest increase from the predicted 10 transits needed to detect an Earth-like atmosphere on TRAPPIST-1~e without considering stellar contamination \citep{Lin_2021}, due to the increase in noise from the spectroscopic ratio.  

Due to the short rotation period of TRAPPIST-1, we restrict our consideration to close transits with less than 8 hours between the transits of TRAPPIST-1~b and e, corresponding to approximately 10\% of the 3.3 day rotation period of TRAPPIST-1 \citep{Luger_2017}. This is a small enough fraction of the rotation period that there should not be a significant amount of stellar surface rotation between observations, but is still flexible enough that we are able to find enough close transit instances in the near future. We observe the entire span of time between the two transits to monitor the star, as this allows us to monitor for flares and also look for rotational modulation in the out of transit flux.  

As the system has significant transit timing variations (TTVs) due to the orbital resonances between the planets, we find our observing windows based on TTV modeling from \citet{Agol_2021}, which has been updated based on the results from the first TRAPPIST-1 JWST programs \cite[including the ones presented here, as well as those published in][and some that have not yet been published]{lim_2023,Radica_2024, Piaulet-Ghorayeb_2025} in \citet{Agol_2024}. We are careful to find transits in which there are no overlapping transits of the other TRAPPIST-1 planets, which happens relatively often and would introduce extra uncertainty into our light curve analysis. Observation windows that contain transits of other planets within the out of transit windows pose no additional complication as we ensure that there is always an hour of non-transit baseline at the beginning and end of the observations, plus extra baseline between the transits. With these considerations, we found the 15 closest transits within JWST Cycles 3 and 4, which come out to about 130 hours of charged observing time. We proposed and were awarded these observations in the JWST Cycle 3 Call for Proposals \citepalias[GO 6456 and 9256 for Cycles 3 and 4, respectively, see][]{t1e_1, t1e_2}.  

%\subsubsection{Caveats}
%(do we even want to get into this in this paper -- no lol)
%\paragraph{What if the stellar surface changes between transits?}
%\paragraph{What if TRAPPIST-1~b has an atmosphere?}
%\paragraph{What if TRAPPIST-1~e has clouds?}
\begin{figure*}
    \centering
    \includegraphics[width=\linewidth]{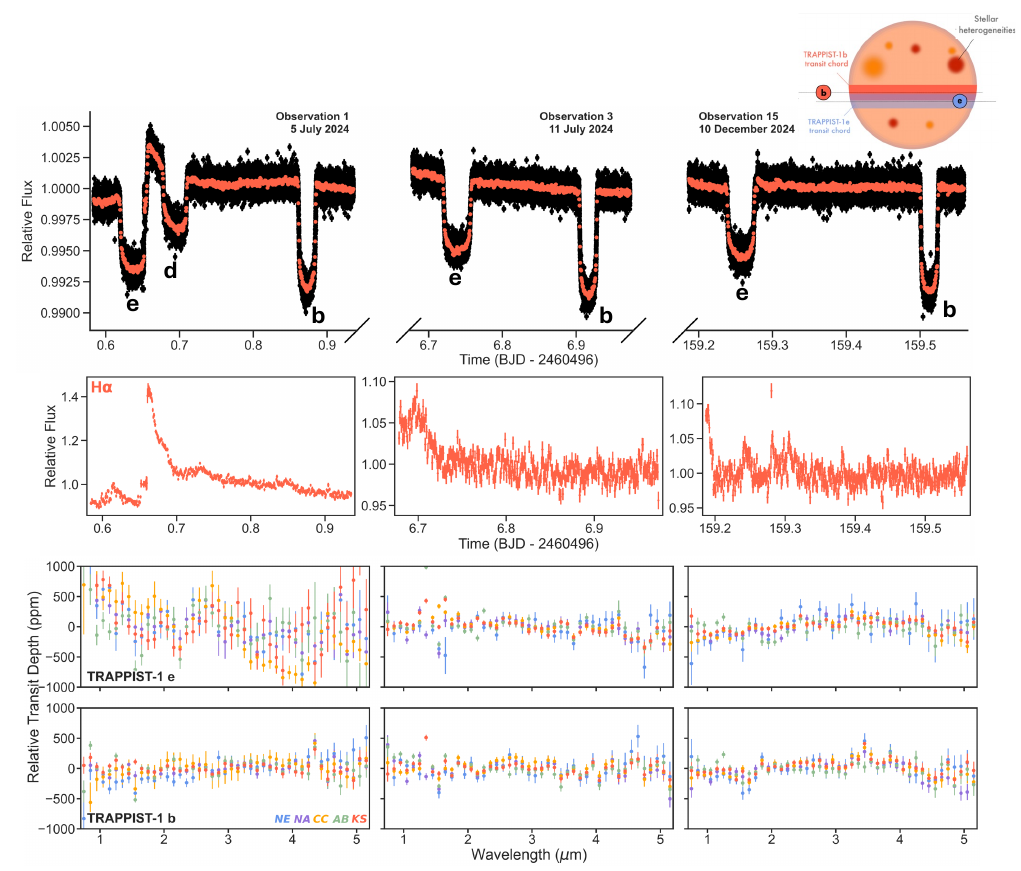}
    \caption{Top: White light curves of observation 1, observation 3, and observation 15 from the \textit{NA} reduction, both at native time resolution in black and binned to 25x in red for visibility. All transits are labeled by planet. Observation 1 also has a large flare that occurs right before the egress of planet e. Shown in the upper right is a schematic of our proposed setup, with the true overlapping planetary transit chords taken from \citet{Agol_2021}. Middle: H$\alpha$ light curve from the \textit{NA} reduction. Peaks in H$\alpha$ correspond to flaring events, many of which are not visible in the white light curve, at least without hints to their location from the H$\alpha$ signature. Note that the light curves are binned in time to 50x (to approximately 70 second bins) relative to that shown in the white light curve for visibility in the low signal single wavelengths shown. Bottom: Transmission spectrum of TRAPPIST-1~e (top) and TRAPPIST-1~b (bottom) for observation 1, observation 3, and observation 15 from left to right. }
    \label{fig:wlc}
\end{figure*}

\section{First Observations}\label{sec:obs}
The first three observations of our multi-cycle JWST program were taken on 5 July 2024 (``observation 1"), 11 July 2024 (``observation 3"), and 10 December 2024 (``observation 15")\footnote{Note the odd numbering of these observations comes from how the observations are designated in the program APT, which are subject to change.}. All observations use NIRSpec/PRISM and the 512 subarray, with 5 groups per integration. Our observations are set up to observe approximately one hour of baseline before the first planet's transit through one hour of baseline after the second planet's transit, the length of which will vary based on the relative alignment of TRAPPIST-1~e and b. This ends up as 22070, 18480, and 23370 integrations and a total science exposure time of 8.4, 7.1, and 8.9 hours for observations 1, 3, and 15, respectively. These observations have a transit of TRAPPIST-1~e and then TRAPPIST-1~b, but observation 1 has an additional transit of TRAPPIST-1~d in between the e and b transits. White light curves of all visits are shown in \autoref{fig:wlc}, top panel. We note that the timing of each of these transits is in good agreement with the original TTV predictions from \citet{Agol_2021}, falling well within the stated uncertainties.

\subsection{Data Reductions}\label{sec:red}
As is standard in analyses of transmission spectra with JWST, we perform multiple independent data reductions with different pipelines to test the reproducibility of our results. All reductions are labeled by the name of the pipeline used, and then by the initials of the co-author who performed the reduction. We describe each reduction below, and summarize them in \autoref{tab:reductions}.

%\begin{deluxetable*}{c|c c c c c}
%    \centering
%    \tablecaption{Quick summary of all data reductions used for analysis.
%    \label{tab:reductions}}
%    \tablehead{\colhead{} &
%    \colhead{NE} & \colhead{NA} & \colhead{CC} & \colhead{AB} & \colhead{KS}}
%    \startdata
%        & \textit{transitspectroscopy} & \textit{transitspectroscopy} &  \textit{ExoTiC-JEDI} & \textit{Juniper} & \textit{Eureka} \\
%         systematics model & linear + GP & linear + GP & GP (1), second-order (3, 15) & third-order (1), second-order (3, 15) & linear (1), second-order (3), linear (15) \\
%         limb darkening & fit $u_1$, $u_2$ & fit $q_1$, $q_2$ & fit $q_1$, $q_2$ & fixed four-parameter non-linear & fixed SPAM coefficients
%    \enddata
%\end{deluxetable*}

\begin{deluxetable*}{c|c c c c}
    \centering
    \tablecaption{Quick summary comparing all of the data reductions used for analysis.
    \label{tab:reductions}}
    \tablehead{\colhead{} &
    \colhead{pipeline} & \colhead{systematics model} & \colhead{limb darkening} & \colhead{transit fitting}}
    \startdata
        NE & \textit{transitspectroscopy} & linear + GP & fit $u_1$, $u_2$ & individual\\
        NA & \textit{transitspectroscopy} & linear + GP & fit $q_1$, $q_2$ & individual (1), joint (3, 15)\\
        CC & \textit{ExoTiC-JEDI} & GP (1), second-order (3, 15) & fit $q_1$, $q_2$ & joint \\
        AB & \textit{Juniper} & third-order (1), second-order (3, 15) & fixed four-parameter non-linear & individual\\ 
        KS & \textit{Eureka} & linear (1, 15), second-order (3) &  fixed quadratic SPAM coefficients & individual (1, 15), joint (3) \\
    \enddata
\end{deluxetable*}

All light curve fits besides reduction \textit{AB} used common planet parameters obtained from a fit to the white light curves for each visit, obtained from the \textit{transitspectroscopy} NE reduction (\autoref{sec:red_ne}), which are given in \autoref{tab:wlc_params}. Reduction \textit{AB} fit for their own white light curve parameters and used that for the rest of their spectroscopic light curve analysis. Spectra from all reductions are shown in \autoref{fig:wlc}, bottom panel. All reductions have been binned to the 100 nm wavelength bins from the \textit{KS} reduction for a direct visual comparison, and are median-subtracted due to some slight offsets due to specifics in the reductions (see discussion in \autoref{sec:interp}), though absolute transit depths for each reduction are shown in \autoref{app:absolute}. Note that the wavelength region from approximately 1.1-1.7 $\mu$m is saturated, which is the cause of the larger outliers present in this part of the spectrum due to different pipeline's handling of saturated pixels. This saturation was deliberate, as saturating in this relatively small wavelength span allows for simultaneous coverage of the rest of the NIRSpec/PRISM wavelength range, which is important for us to diagnose the stellar contamination. A more thorough analysis of the recovery of this saturated region will be the topic of a future work.

\begin{deluxetable*}{c c c c}
    \tablecaption{Orbital parameters used for spectroscopic light curve analysis. Orbital parameters for TRAPPIST-1~e were fixed to those from \citet{Espinoza:2025}, while the others were obtained from our new transits. \label{tab:wlc_params}}
    \tablehead{\colhead{} &
    \colhead{TRAPPIST-1~e} & \colhead{TRAPPIST-1~d} & \colhead{TRAPPIST-1~b}}
    \startdata
    period (days) & $6.101012$ & $4.049219\pm0.000026$ & $1.510826\pm0.000006$  \\
    a/R$_*$ & $52.72$ & $40.39^{+0.14}_{-0.18}$ & $20.81^{+0.04}_{-0.06}$ \\
    impact parameter (b) & $0.172$ & $0.068^{+0.054}_{-0.064}$ & $0.069\pm0.030$ \\
    %inclination (i, $^{\circ}$) & $89.81$ & 89.903 & 89.809 \\
    T0 July 5, 2024 (BJD TDB) & $2460496.637930\pm0.000044$ & $2460496.693679\pm0.000023$ & $2460496.872697\pm0.000011$ \\
    T0 July 11, 2024 (BJD TDB) & $2460502.738577\pm0.000015$ & -  & $2460502.916128\pm0.000016$ \\
    T0 December 10, 2024 (BJD TDB) & $2460655.257098\pm0.000045$ & -  & $2460655.512395\pm0.000015$ 
    \enddata
\end{deluxetable*}

\subsubsection{\textit{transitspectroscopy}, NE}\label{sec:red_ne}
This reduction utilizes the \textit{transitspectroscopy} data reduction pipeline \citep{transitspectroscopy}, whose detailed explanation is given in \citet{Espinoza:2025}. In short, this makes use of the \textit{JWST} Calibration Pipeline \citep{jwst-calibration-pipeline} version 1.12.5, with a few modifications: no dark current step is used, we perform our own 1/f and background corrections, we use our own saturation reference file which has a threshold 90\% that of the default reference file for this subarray and we perform jump detection through differences of groups via a time-series on all integrations; the details of those techniques are described in detail in \cite{Espinoza:2025}. An aperture with radius of 7 pixels is used for spectral extraction.

For the transit fits, each planet was fit individually using \textit{juliet} \citep{juliet}. Following \cite{coulombe:2024}, we fit for limb-darkening but using the $u_1$ and $u_2$ parametrization instead of the \citet{KippingLDs} parametrization, using uniform priors between -3 and 3 for each. We fix the period, time of transit center and orbital parameters to the ones in Table \ref{tab:wlc_params}. We set a uniform prior for $R_p/R_s$ between 0 and 0.2, a normal prior for the \texttt{mflux} (relative flux offset) parameter with mean 0 and standard deviation 0.1, a log-uniform prior between 0.1 and 30,000 ppm for an additive jitter component to the errorbars, and fit for two systematic terms: a slope (which uses the standardized times as regressors) with a uniform prior between -10 and 10, and a gaussian process (GP) with a Mat\`ern 3/2 kernel, with a log-uniform prior on its amplitude of 1 to 10,000 ppm. This is the same systematics model used for previous observations of TRAPPIST-1~e \citep{Espinoza:2025}, and fits the long term instrumental trend with the slope and the short frequency time variability with the GP. For Observation 15, the timescale of the process was fixed to the one found in the white-light light curves, which was 0.71 hours, for the spectroscopic light curve fits, while for the other observations the timescale was fitted with a log-uniform prior on the timescale between $10^{-5}$ and $10^3$ days. The lightcurves were binned before fitting in time to a 10-point (approximately 14 second) average, and only the local regions in time within 2 hours of mid-transit were used to fit the transits --- the only exception being Observation 1, where the cut was made  for e just before the big flare.

\subsubsection{\textit{transitspectroscopy}, NA}\label{sec:red_na}
This reduction utilizes the \textit{transitspectroscopy} data reduction pipeline. This reduction closely follows the NE reduction, but for the spectral extraction, an optimal extraction with an aperture of 12 pixels was used. In observation 1, each transit was fit individually to avoid the peak of the flare, using integrations 0 to 3800 for TRAPPIST-1~e, 5000 to 10000 for TRAPPIST-1~d, and 10000 to the end (22070) for TRAPPIST-1~b. For observation 3 and 15, both TRAPPIST-1~e and b were fit simultaneously. Spectroscopic light curves are fit with \textit{dynesty} \citep{dynesty} at the native resolution using \textit{juliet}, with sampled parameters $R_p/R_s$ (for each planet with uniform prior from 0 to 0.1), additive error scaling factor (with log-uniform prior from 0.1 to 1000), $q_1$ and $q_2$ quadratic limb darkening parameters from \citet{KippingLDs} (using a non-informative 0 to 1 uniform prior), and a systematics model. For all observations, an approximate Matern-3/2 GP kernel from \textit{celerite} \citep{celerite} (with wide log-uniform priors from $10^{-6}$ to $10^2$ on all parameters) and a linear model (with uniform prior between -10 and 10), both with time as a regressor, are used as the systematics model. This systematics model was chosen through a comparison of different systematics models in fitting the white light curve, for which this combination of models was preferred by the resulting Bayesian evidence.

\subsubsection{\textit{ExoTiC-JEDI}, CC}\label{sec:red_cc}
This reduction utilizes the \textit{ExoTiC-JEDI} \citep{exoticjedi} data reduction pipeline. The default \textit{JWST} pipeline v1.12.5 is used to perform the Stage 1 calibration, but the jump detection and superbias steps are skipped. The \textit{ExoTiC-JEDI} custom destriping step is used to correct for 1/f noise, and then the standard ramp-fitting algorithm from the \textit{JWST} pipeline is used to get the rates per integration, but with all groups that reached $>80$\% saturation masked out.

To trace the spectra, a Gaussian was used to find the peak, and x-pixels from 50 to 481 were used. Background extraction is done per column, with the pixels used for background set as $10\times$ the FWHM of 1.1 pixels from the aperture used for the spectral extraction. For the spectral extraction, optimal extraction was used with an aperture of $3\times$ the FWHM. 

For all observations, all transits are fit together and at native resolution. Spectroscopic light curves are modeled with \textit{batman} and
fit with \textit{dynesty}, with sampled parameters $R_p/R_s$ (for each planet with uniform prior from 0 to 0.5), multiplicative error scaling factor (with uniform prior from 0 to 5), $q_1$ and $q_2$ quadratic limb darkening parameters from \citet{KippingLDs} (using a non-informative 0 to 1 uniform prior), and a systematics model, with all other values fixed to those listed in \autoref{tab:wlc_params}. The systematics from observation 1 are fit using the \textit{celerite} quasi-periodic GP kernel to deal with the flare, while observation 3 and 15 are fit with a quadratic polynomial function, all with time as the regressor. The more complex GP systematics model was chosen for observation 1 due to the large flare, while the other two observations were instead fit with the simpler polynomial function as has been found to match NIRSpec systematics from previous observations \citep[e.g.][]{Espinoza:2025}.

\subsubsection{\textit{Juniper}, AB}\label{sec:red_ab}
This reduction utilizes the \textit{Juniper} data reduction pipeline. The default \textit{JWST} pipeline v1.15.1 is used to perform the Stage 1 calibration, utilizing the saturation, refpix, linearity, dark current, and superbias correction steps, but skipping the jump detection step. Background subtraction is then done at the group-level to remove 1/f noise, and then the standard ramp-fitting algorithm from the \textit{JWST} pipeline is used to get the rates per integration.

To trace the spectrum, the \textit{JWST} pipeline Spec2Pipeline step is used, truncating the array to 431 pixels from 0.55-5.38 $\mu$m. Cosmic rays are then flagged iteratively at $6\sigma$, while the \textit{JWST} pipeline flagged pixels are masked. An additional 1/f subtraction is then performed using the first 3 and last 2 rows on the detector as background. A standard spectral extraction is then performed with a 4.5-pixel half width around the trace.

The white light and spectroscopic light curves are fit using \textit{emcee} \citep{emcee} and the \textit{batman} \citep{batman} transit model, with the P, t0, a/r$_*$, and inclination from the white light curve fit then fixed in the spectroscopic light curves, and all light curves fitting for $R_p/R_s$ (with uniform prior from 0.0001 to 0.2). A zero eccentricity orbit is assumed for all planets, and spectroscopic light curves are fit at the native resolution. Each transit for each visit was fit individually. For observation 1, the e transit spanned integrations 200-4253 (dropped first 200 integrations of exposure to avoid ramp effects), the d transit spanned integrations 4800-10000, the b transit spanned integrations 13786-22069, and a third order polynomial is used for systematics detrending due to the complexity presented by the flare and its decay. For observation 3, the e transit spanned integrations 0-9240, the b transit spanned integrations 9240-18479, and a second order polynomial is used for systematics detrending. For observation 15, the e transit spanned integrations 1000-5750, the b transit spanned integrations 18000-23369, and a second order polynomial is used for systematics detrending. The higher order polynomial for systematics used in observation 1 was not found to be necessary for observations 3 and 15 since there were no obvious flare or stellar activity features included in the integrations used for the transit fits. The limb darkening is fixed to values found with ExoTiC-LD \citep[database 3.1.2,][]{Grant2024}, by using trilinear interpolation between the most TRAPPIST-1-like PHOENIX models. The four-parameter non-linear limb darkening law is used. 

\subsubsection{\textit{Eureka}, KS}\label{sec:red_ks}
This reduction utilizes the \textit{Eureka} \citep{eureka} data reduction pipeline v1.2. We use the default \textit{JWST} pipeline v1.15.1 using the CRDS context 1303. We then perform group level background subtraction with a 4-sigma rejection threshold, and also do jump step rejection with a 4-sigma rejection, before using the standard ramp-fitting algorithm to get the rates per integration. 

To determine the source position, we compute the median flux along each row (i.e., along the dispersion direction) spanning detector pixel columns 60 to 460 and then fit a Gaussian to the resulting 1D profile. For optimal spectral extraction, we use a 7 pixel full width aperture and the median integration frame as the weighting.

The spectroscopic light curves are fit using \textit{emcee} with a \textit{batman} \citep{batman} implementation of the \citet{Mandel2002} transit model. Quadratic limb darkening is used, with values fixed to limb darkening coefficients determined using the MC-SPAM algorithm as discussed by \cite{ld} and previously used by \citet{Espinoza:2025}. 
For observation 1, we fit each planet transit individually. For the e transit, we use integrations 50 -- 1800 for the baseline and 2900 -- 4150 for the in-transit region.  We use a linear function in time to detrend each spectroscopic channel.  Without any post-transit baseline, a more sophisticated model would yield spurious transit depths.  We did not fit for the transit of planet d.  For planet b, we used integrations 12000 -- 21770 and a quadratic function in time for detrending.  Shortward of 2 $\mu$m there is clear evidence of curvature in time.  For observation 3, we fit the transits of planets b and e simultaneously.  We trim the first 300 integrations in time and, due to the visually apparent curvature, fit the remaining baseline using a quadratic function.
Similar to observation 1, we separately fit the two transits from observation 15.  We use the first 7000 integrations to fit planet e and the last 8370 integrations to fit planet b. Due to the lack of apparent curvature, both use a linear function in time for detrending. We fit 46 spectroscopic light curves spanning 0.7-5.3 $\mu$m at constant resolution (100 nm). All other values are fixed to white light curves values from \autoref{tab:wlc_params} besides $R_p/R_s$. In all cases, we apply a uniform prior from 0.02 to 0.15.

%\begin{figure*}
%    \centering
%    \includegraphics[width=\linewidth]{observation_compare_ksgrid.png}
%    \caption{Transmission spectrum of TRAPPIST-1~e (top) and TRAPPIST-1~b (bottom) for observation 1, observation 3, and observation 15 from left to right. }
%    \label{fig:spec}
%\end{figure*}

\subsection{Interpretation}\label{sec:interp}
We present a brief description of the light curve morphology for each of our three observations.
\subsubsection{Observation 1}
Just before the egress of the transit of TRAPPIST-1~e in observation 1, a large flare appears. This significantly complicates this observation, especially since, looking at wavelength covering H$\alpha$, it seems the flare decay lasts through the entire rest of the time series (see \autoref{fig:wlc}, middle panel).%, and inherently breaks the assumption in our proposed method of stellar contamination transfer that the stellar surface does not appreciably change between the transits of e and b. 

The alignment of our planet transits in this observation allows us an interesting probe of the spectroscopic nature of the stellar surface pre- (transit e), during (transit d), and post-flare (transit b). Comparing the transmission spectra of planets e and b suggests that the flare may have caused a lasting change in the stellar flux. Although the spectrum of planet e has significantly larger uncertainties than expected due to the transit egress being hidden by the flare, there still seems to be an upward slope into the optical indicative of the presence of cold surface features, as well as some bumps through the spectrum that look to be due to H$_2$O, though our analyses are not in complete agreement longwards of 3 $\mu$m. On the other hand, the spectrum of planet b shows a downwards slope into the optical, instead indicative of a hot dominant surface component, and appears relatively flatter throughout the rest of the wavelength range. This change in flux before and after the flare can be seen in the baseline as well, as the post-flare baseline is noticeably higher than pre-flare, even after what seems to be the full decay of the flare. This is similar to the long-term ($\sim$day timescale) post-flare brightening \citet{Morris_2018} noted in the K2 photometric monitoring of TRAPPIST-1. This may signal a longer-lived hot surface component after the flare, perhaps indicative of heating caused by material falling back down to the surface post-flare, like Partially Erupted Prominence Material on the Sun \citep[e.g.][]{Gilbert_2013}.

The transmission spectrum of planet d is more complex to analyze, as it significantly changes as a result of the detrending method used in the light curve fitting due to the sharp change in the baseline through the transit due to the flare. We believe that an in-depth analysis of these observations relies on the use of simultaneous flare and planet modeling during the light curve fitting step, which we leave to a future work. We show a few illustrative figures of TRAPPIST-1~d in \autoref{app:d} for completeness.

\subsubsection{Observation 3}
Observation 3 appears clean in comparison to observation 1, with no pronounced flares seen in the white light curve. There is a clear downwards slope in the time series, but this is at least partially due to the known systematic effect on the NRS1 detector, which the NIRSpec/PRISM instrument illuminates \citep[see e.g.][]{Espinoza_2023, moran_2023, Rustamkulov_2023}. There are some bumps present in the resulting TRAPPIST-1~e and b transmission spectra, but with an overall relatively small amplitude.

\begin{figure*}
    \centering
    \includegraphics[width=0.9\linewidth]{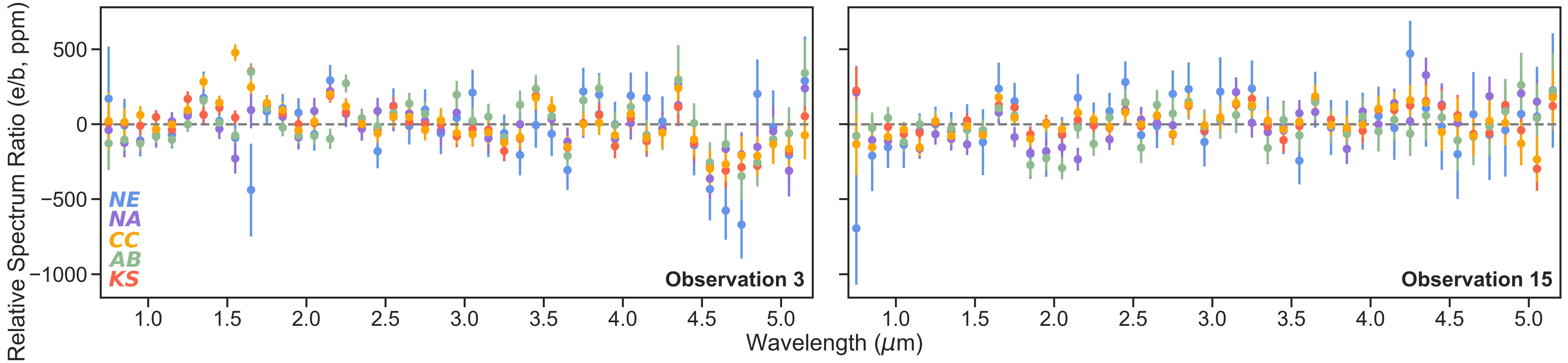}
    \caption{Relative transmission spectrum ratio of TRAPPIST-1~e with TRAPPIST-1~b, for observation 3 (left) and observation 15 (right). The relative transmission spectrum ratio for observation 1 is not shown due to the large error bars for the TRAPPIST-1~e spectrum from this observation due to the large flare. The multiple reductions show close agreement for the most part, though the difference in estimated error bar size due to a combination of number of free parameters, method of systematics removal, and included integrations for each reduction can be seen.}
    \label{fig:ratio}
\end{figure*}

\subsubsection{Observation 15}
Observation 15 once again appears quite simple upon a first look at the white light curve, but there is a small flare visible in the white light curve right after the egress of the transit of planet e. Also, there seems to be no clear downward slope like that seen in observation 3. This may be due to rotational modulation from active regions rotating on and off the stellar surface over the course of the transit, causing a signal opposite to the instrumental slope and therefore canceling it out. %This interpretation is bolstered by looking at specific high signal wavelength light curves, such as the continuum light curve in \autoref{fig:flares}, in which an upwards slope can be seen through the observation. 
The instrumental slope always creates a downwards signal \citep[see e.g.,][]{Espinoza_2023}, such that this upwards slope is evidence for an astrophysical origin.

The spectra of planets b and e both look to be dominated by contamination from a hotter surface component due to their downwards slopes into the shorter optical wavelengths. Overall, the spectrum of planet e looks to have a large curvature to it, sloping upwards to around 3 $\mu$m and then curving back down into the longer wavelengths, with some additional small bumps throughout. The spectrum of planet b is similar, though with potentially some sharper features throughout the spectrum -- most notably around 3.5 $\mu$m and 1.7 $\mu$m, though the latter is on the edge of the saturated region, making its origin less clear. Otherwise, the curvature of the spectrum matches, sloping upwards to around 3 $\mu$m and then falling back down.

%\begin{figure*}
%    \centering
%    \includegraphics[width=\linewidth]{halpha_comp_full.png}
%    \caption{Comparison between H$\alpha$ (top) and the continuum (bottom) for each of our three observations. Peaks in H$\alpha$ correspond to flaring events, many of which are not visible in the continuum, at least without hints to their location from the H$\alpha$ signature. Note that the light curves are binned in time to 50x (to approximately 70 second bins) relative to that shown in \autoref{fig:wlc} for visibility in the low signal single wavelengths shown.}
%    \label{fig:flares}
%\end{figure*}

\section{Discussion}\label{sec:discussion} 

\subsection{The use of close transits for stellar contamination correction}
Each of the first three observations essentially represents a different scenario for the use of close transit observations. We discuss each independently before we make a comparison. We show the relative spectrum ratio of observations 3 and 15 for all of our reductions in \autoref{fig:ratio}. Note we do not show the relative spectrum ratio for observation 1 due to the large error bars on the TRAPPIST-1~e spectrum because of the flare. For a single observation, the signals from an atmosphere on TRAPPIST-1~e would be below the noise. Therefore, if the method works as theorized, we expect to see no significant deviations from a flat line in the ratio spectra. %We mainly focus on wavelengths of $2\mu$m and greater since according to our simulations we gain all of our atmospheric information in this region, and because the $\sim 1-2 \mu$m region suffers from saturation.

\subsubsection{Large disruptive flare in observation 1}
Our observations strongly suggest that the assumption that the stellar surface is consistent between the transits of b and e is clearly broken in observation 1 due to the large flare: TRAPPIST-1~e's spectrum contamination suggests, due to the upward slope into the optical, a surface dominated by unocculted cold spots, while TRAPPIST-1~b's downward slope suggests one dominated by hot spots. The ability of flares to impart lasting changes to the surface heterogeneities of stars was recently suggested by analyses of other observations of TRAPPIST-1 as well \citep{Vasilyev_2025}. In addition, the size of the errors on the transmission spectrum of TRAPPIST-1~e from this observation are significantly higher than expected, due to the flare disrupting the observation before the transit ended, leaving uncertainty in the shape of the transit egress when fitting the light curves. This observation essentially represents a worst-case scenario for the use of close transit observations, since the flare both impacted the transit fit itself, and changed the surface of the star between the observations. We leave the full analysis of this observation for future work that will account for the transits and the flare simultaneously.

\subsubsection{Deviation from a flat line in observation 3}\label{sec:obs3_disc}
\begin{figure*}
    \centering
    \includegraphics[width=0.5\linewidth]{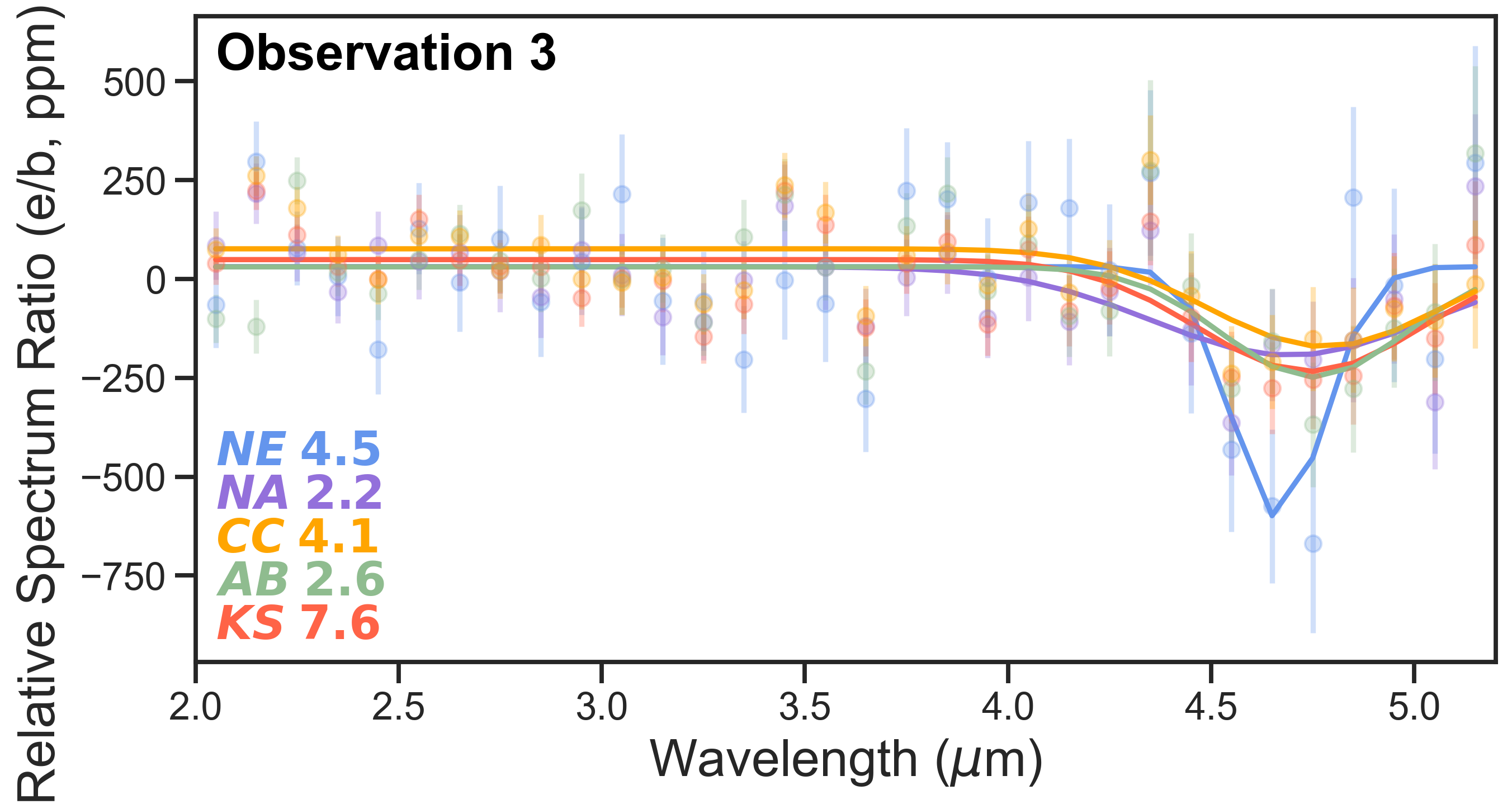}
    \includegraphics[width=0.3\linewidth]{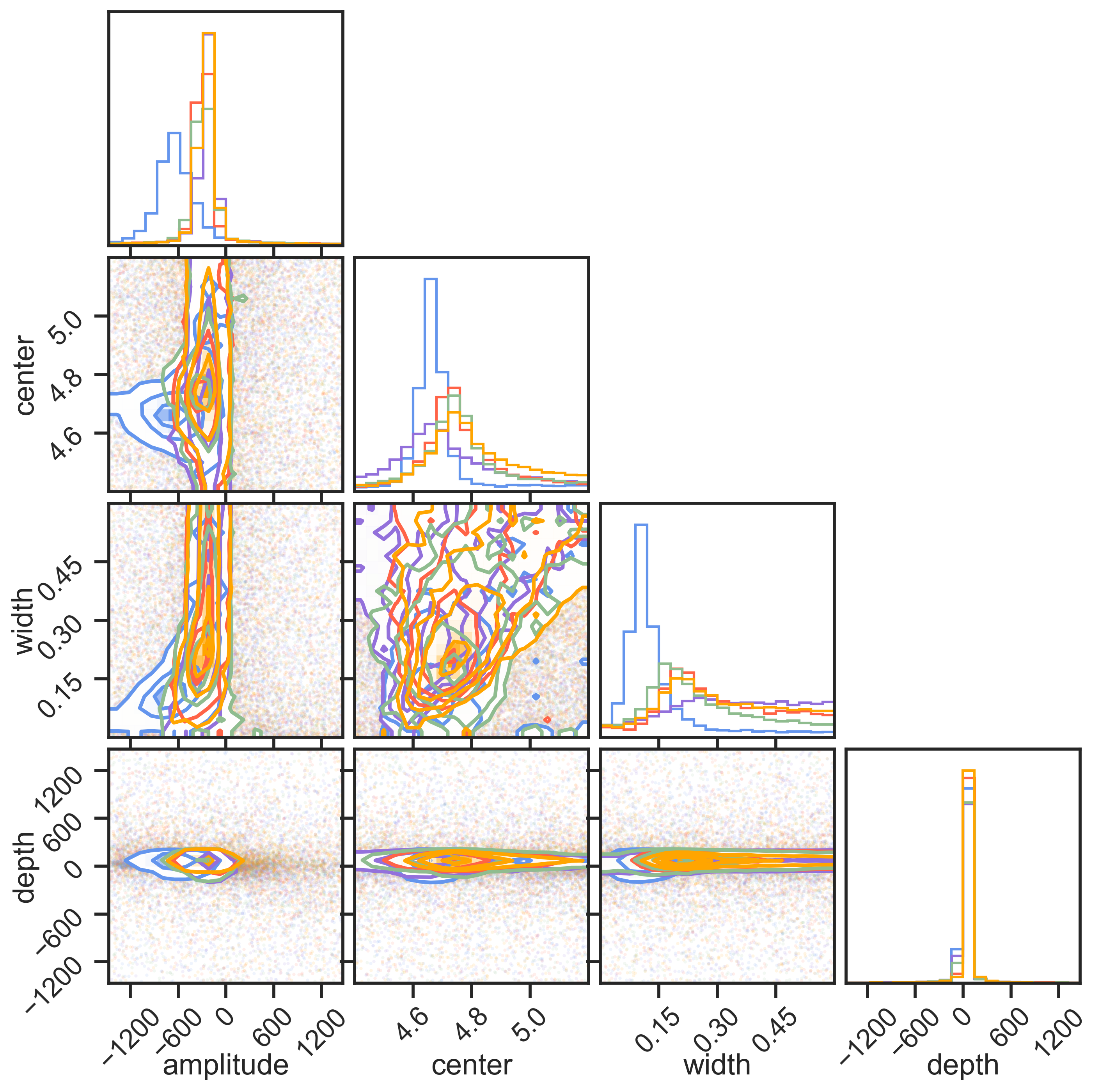}
    \caption{Gaussian feature fits to observation 3's relative spectrum ratio. Given next to each reduction is the $\Delta$ ln Z preference of the Gaussian feature over a flat line, which vary from weak to strong evidence depending on the reduction. Also shown are the posteriors for the parameters fit for the Gaussian feature, showing general agreement between reductions.}
    \label{fig:gauss}
\end{figure*}

Between 4 and 5 $\mu$m in observation 3, there does appear to be a bump that deviates from the expected flat line from the full removal of stellar contamination, which is seen as a positive feature in the TRAPPIST-1~b spectrum and a negative feature in the TRAPPIST-1~e spectrum (see \autoref{fig:wlc}, bottom panel). To test the significance of this feature, we fit a Gaussian feature of the form $A\,\,e^{-(x-\mu)^2/(2\sigma^2)} + D$ to the spectrum ratio, fitting for amplitude $A$ (uniform from -0.2 to 0.2), peak location $\mu$ (uniform from 4.4 $\mu$m to 5.2 $\mu$m), Gaussian width $\sigma$ (uniform from 0 to 0.6), and flat line transit depth $D$ (uniform from -0.2 to 0.2) using \textit{dynesty} \citep{dynesty}. We only consider wavelengths above 2 $\mu$m, since the saturated region in this observation looks to be especially affected by outliers. We compare the resulting Bayesian evidence to that of a flat line, and we find that all of the reductions prefer the Gaussian feature to a flat line to varying degrees of confidence. These fits are shown in \autoref{fig:gauss}, with the $\Delta$ ln Z value between the Gaussian feature and a flat line given beside each reduction name, varying from weak to strong evidence \citep{Trotta:2008}. In addition, the right side of \autoref{fig:gauss} shows the posteriors for each fit, for which all Gaussian fits agree that the peak is somewhere around $4.7\mu$m. 

We first attempt to find an instrumental explanation for this feature, considering the following aspects.
\paragraph{Background subtraction} Perhaps the background as a function of wavelength could vary in such a way over the observation that it induces a feature in this location. We calculate the time series of the average background per column and find it to be flat and non-varying in both wavelength and time, meaning it could not impart this feature.
\paragraph{PSF/FWHM} If the JWST PSF was changing significantly over the course of the observation, that may be able to cause a false feature. We calculate the FWHM of each column of the extracted spectrum and the resulting time series as a function of wavelength. We also look at the guide star observations and its FWHM as a function of time using \textit{spelunker} \citep{spelunker}, which will show if something like a mirror tilt event or guiding anomaly happened with the telescope during the observation. In both time series, we see a small slope in the FWHM in approximately the first hour of the observation, as well as a small jump in the time series for the guide star observations. However, this seems to impart a relatively small effect on the resulting flux which our detrending methods are able to deal with, and importantly does not appear to have any structure with wavelength, and is therefore unlikely to cause this feature.
\paragraph{Spectral trace variations/slit losses} A significant variation in the location of the spectrum on the detector may have an effect on the resulting transmission spectrum. We take the time series of the spectral trace location as calculated during our data analysis process, but find it to vary by only a fraction of a pixel across the detector over the course of the time series, and therefore is unlikely to be the source of this feature.
\paragraph{Aperture for spectral extraction} We test spectral extraction using apertures from 2 to 15 pixels and find that this feature is robust to the choice of aperture size. This is in agreement with the feature's presence in all reductions, which use many different apertures as detailed in \autoref{sec:red}, and therefore the aperture size is not affecting the presence of this feature.
\paragraph{Individual pixel light curves}
If individual pixels, especially ones near the center of the trace, were behaving oddly, that may be able to imprint a signature in the spectrum in time. Therefore, we look at the individual illuminated pixel light curves between 4 and 5 $\mu$m, but find no odd behavior and therefore rule this out as a likely cause of the feature.
\paragraph{Instrumental systematics} As previously mentioned in \autoref{sec:obs}, there is a known instrumental slope in the NRS1 detector, which is the detector used for all NIRSpec/PRISM observations \citep[also see][]{Espinoza_2023}. To date, there are no published efforts to characterize the specifics of this slope. In addition, observations of TRAPPIST-1 pose an additional complication, since the rapid variations in the stellar spectrum can cause astrophysical changes to the stellar baseline in time, even over the relatively short timeframe of a single transit observation \citep[see e.g.][]{Rathcke_2025}. It is beyond the scope of this paper to directly characterize the instrumental slope, though we do an analysis of the change in stellar spectrum throughout the observation in \autoref{app:sys} and find the slope is unlikely to be the cause of the feature. \\ %In addition, the fact that the same feature is not present in either of the other two observations, taken with the same observing mode of the same target, makes it more unlikely to be instrumental in nature.

As none of these investigations lead conclusively to this feature, we consider an astrophysical explanation. The lack of this same signature (a relatively high transit depth on TRAPPIST-1~b and low transit depth on TRAPPIST-1~e around 4.7 $\mu$m) in observation 15 makes it less likely that this feature has any connection to the planets themselves, so we consider the star. One possible explanation a strong CO feature present at these wavelengths (spanning from 4.3 to 5.1 $\mu$m), which is known to be present in M dwarf stars. The strength of CO emission on the Sun is affected by activity, and can vary significantly on relatively short ($\sim$hour) timescales \citep[e.g.][]{Stauffer_2022}. In magnetohydrodynamic (MHD) stellar simulations, CO is a sensitive temperature probe \citep{Norris_2023}. However, we are uncertain if a similar effect would be present on a star like TRAPPIST-1, nor the strength of said effect. Recent MHD models of spots on M0 stars show a significant opacity contribution between 4-5 $\mu$m due to diatomic molecules, including CO, that is missing in radiative equilibrium photospheric models \citep{Smitha_2025}, such that CO could be present in either changing surface features or potentially tied to flaring behavior. %However, a dip in the spectrum at these same wavelengths does seem to be visible in observation 15 for both planets (see \autoref{fig:spec}), which may again point to a stellar origin. 

To further investigate the potential for activity to be causing this contamination, we look at the H$\alpha$ wavelength, which is especially sensitive to flares, shown in \autoref{fig:wlc}, middle panel. We do see evidence for a relatively small flare that appears just before the transit of planet e, and in addition the H$\alpha$ flux is elevated at the beginning of the observation. This may point to the presence of a larger flare that appeared before the beginning of the observation, for which we are still seeing the decay. This is not unexpected since \citet{Howard_2023} predict $\sim3.6$ flares of $10^{30}$ erg in the TESS bandpass per day and this observation is seven hours long. If this is the case, the flare decay and the temperature variability it creates may be tied to the presence of the variability seen between 4 and 5 $\mu$m. This is by no means a definitive explanation -- further investigation in future observations, as well as more sophisticated models of M dwarf active regions, are needed for that. However, the potential for stellar contamination at these wavelengths is an important consideration for future study, as this could complicate the detection of the 4.3 $\mu$m CO$_2$ feature that we are reliant on for atmospheric detection. Besides this potential feature at 4.7$\mu$m, the rest of the spectrum ratio (outside of the saturated region) is consistent with a flat line, though there was not as clear of a stellar contamination signal in this observation as in the other two observations overall.

\subsubsection{Dependence on reduction in observation 15}\label{sec:obs15_disc}
This observation looks to be the best chance to test how well the ratio of transmission spectra to remove the stellar contamination works among our first three visits, as there are strong signs of stellar contamination signals present in both the TRAPPIST-1~b and e spectra, and the spectrum ratio in \autoref{fig:ratio} does not have any obvious deviations from a flat line as is seen for observation 3. This provides a clear opportunity to evaluate the effectiveness of the spectral ratio method. We perform a series of tests to attempt to measure the improvement in the contamination signal due to the spectrum ratio.

We first attempt to test if the spectrum ratio is more closely described by a flat line than the initial spectrum of planet e (i.e., if the ratio makes the spectrum ``flatter", or less affected by stellar contamination, than the original spectrum was). One way to do this is with a simple $\chi^2$ test before and after the spectrum ratio is applied, but since taking the spectrum ratio results in uncertainties larger than in the original spectra of TRAPPIST-1~e, it is not suitable for our case (see \autoref{sec:sims}).  %We calculate a simple $\chi^2$ against a flat line for each reduction of the spectrum of planet e, as well as the spectrum ratio with planet b. In every reduction, the $\chi^2$ value is smaller for the spectrum ratio than it is for the individual spectrum of planet e. However, this analysis is complicated by the fact that the error on the spectrum ratio (described by \autoref{eq:err}) is relatively larger than that in the original spectrum of e, since it takes into account the error from multiple measurements (planets b and e). This can be seen by eye through the relative error bar size between the left and right sides of \autoref{fig:gps}. This means there is some inherent lowering of the $\chi^2$ associated with the spectrum ratio, since the uncertainty on the ratio measurement is larger than the individual spectra. As an example of this, even with the deviation from a flat line present in observation 3, the $\chi^2$ of the spectrum ratio is lower than that of the individual spectrum for most of the reductions.

We instead test the removal of stellar contamination through the use of GPs. Since we have no accurate forward models for the stellar contamination \citep[see efforts in \autoref{app:ret_fits}, and examples from][]{Espinoza:2025}, GPs allow us to model-independently estimate the ``feature size" present in the spectrum through the amplitude of the GP. Note that we are not removing the stellar contamination with this GP model, but using it as a proxy for deviations from a flat line, which should be dominated by the stellar contamination signal for any single visit. We use an approximate Matern-3/2 kernel from \textit{celerite} \citep{celerite} for our GP model.

We fit the original TRAPPIST-1~e spectrum and the ``corrected" TRAPPIST-1~e spectrum from each reduction. To get the corrected spectrum, we take the ratio of the TRAPPIST-1~e and b spectrum, and then multiply by the median transit depth of TRAPPIST-1~b to renormalize the spectrum to be on the same scale. This enables a direct comparison in feature amplitude between the two spectra. We then compare the evidence from this GP fit to a flat line, before and after taking the ratio, to see if the spectrum is more consistent with a flat line after taking the ratio.

\begin{figure*}
    \centering
    \includegraphics[width=\linewidth]{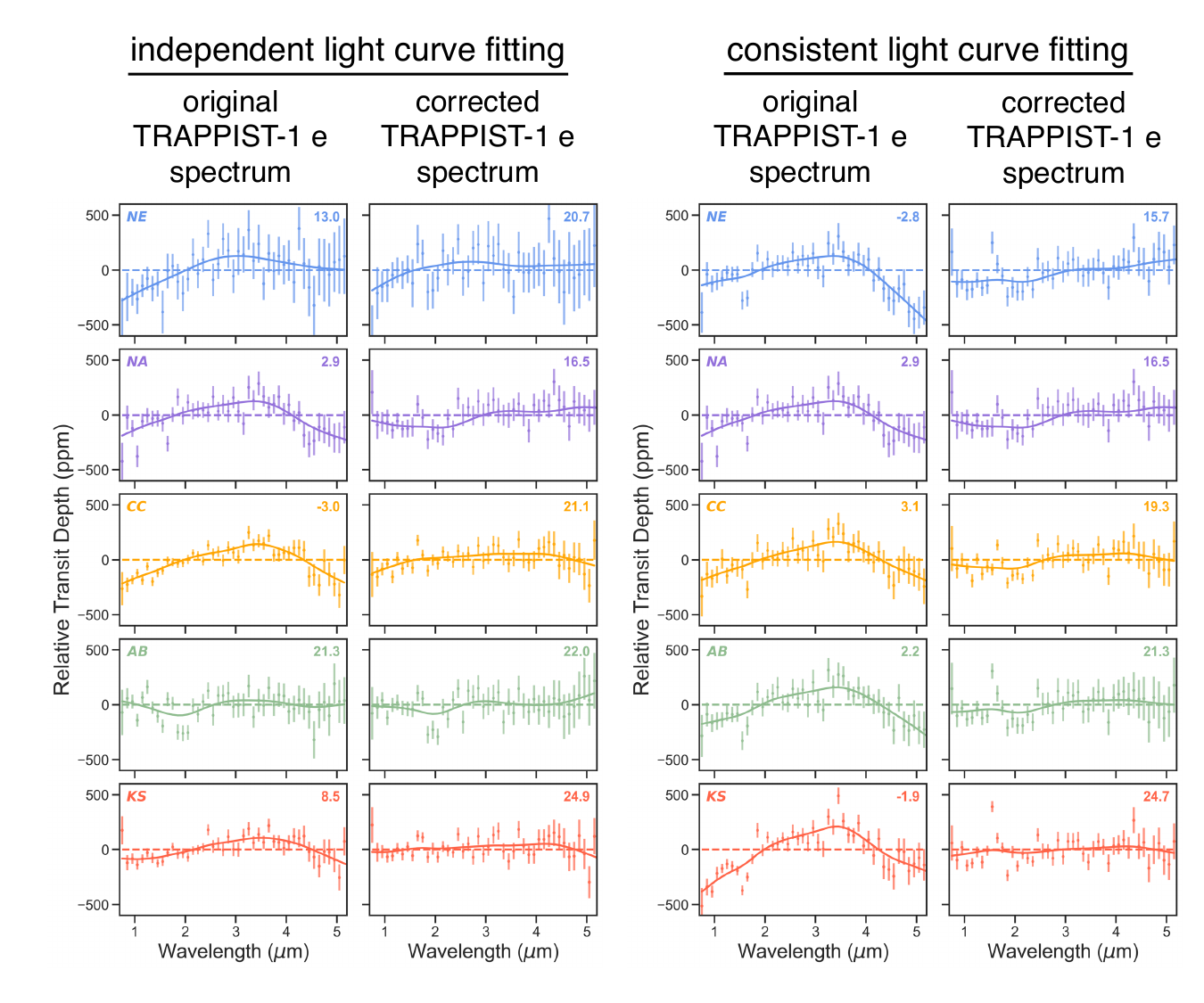}
    \caption{The original TRAPPIST-1~e spectrum and the corrected TRAPPIST-1~e spectrum for observation 15, where ``corrected" means dividing by the TRAPPIST-1~b spectrum and then multiplying by the median transit depth of the TRAPPIST-1~b spectrum to return them to the same scale. Plotted on top are GP and flat line fits to the data. The number in the top right corner is the $\Delta$ ln Z between the flat line and GP models, such that a negative number prefers a GP and a positive number prefers a flat line. On the left are the results from the independent light curve fitting described in \autoref{sec:red}, and on the right are the results from the consistent light curve fitting test carried out using the setup from the \textit{NA} reduction. For all reductions and both cases, the corrected spectrum more strongly prefers a flat line over the GP model than the original spectrum, which suggests that the corrected spectra are more consistent with a flat line and are corrected, at least in part, for the stellar contamination signal. The shape of the GP model for the corrected spectra differs somewhat for each reduction in the independent light curve fitting case, but the shape looks very similar across the reductions for the consistent light curve fitting case. For all cases, a flat line is always more strongly preferred after the spectroscopic ratio is taken.}
    \label{fig:gps}
\end{figure*}

\autoref{fig:gps} (left panel) shows the results. For all reductions, the corrected TRAPPIST-1~e spectrum more strongly favors a flat line over the GP model than the original TRAPPIST-1~e spectrum (see values in \autoref{fig:gps}), which suggests that the corrected spectrum is flatter and therefore we are removing at least in part the stellar contamination component. However, as can be seen in the left panel, right column of \autoref{fig:gps}, each reduction has a somewhat different shape to its ``residuals" in the corrected spectrum, shown by the shape of the GP model. This is essentially equivalent to the problem with atmospheric retrievals picking up details in the spectra that may seem insignificant by eye -- even though the reductions look by eye consistent across the board, the overall specifics of each reduction have their own unique qualities across the wavelength range as a whole. These small differences are then amplified with the spectroscopic ratio, since this includes all of the details from both planet spectrum fits. 

The treatment of flares in the reductions illustrates this sensitivity. Reductions \textit{NE} and \textit{AB} only consider the integrations before the flare right after the end of the TRAPPIST-1~e transit (clearly visible in H$\alpha$ in \autoref{fig:wlc}, middle panel), reduction \textit{NA} fits it with a GP, while reductions \textit{CC} and \textit{KS} do not explicitly fit for the flare. These small differences in reductions, not even clear by eye, have important implications for the further characterization of the potential atmospheric signal of TRAPPIST-1~e, especially when compounded by the need to stack many visits together. Once again, our analysis is made more difficult because of the presence of flares, especially since we do not have good flare models for this wavelength range. However, regardless of their differences, all corrected TRAPPIST-1~e spectra prefer to be fit by a flat line over the GP model with stronger evidence than in the original TRAPPIST-1~e spectra, to varying degrees of significance (see values in \autoref{fig:gps}). 

While this difference in reduction may not be significant at this point in time, our final result will need to confidently combine many observations, so we begin to investigate the source of discrepancies per pipeline at this point. To identify which part of obtaining the transmission spectrum creates the most significant difference in the final spectrum due to the different reduction, we perform a test where we take each reduction's spectroscopic light curves and fit them in the same way, using the method for the \textit{NA} reduction described in \autoref{sec:red_na}. The results of this test are shown in \autoref{fig:gps}, in the right panel. The spectra from these fits are shown in \autoref{app:fitting_test}. In these results, it can be seen that the original and corrected TRAPPIST-1~e spectra, when compared between reductions, are very similar. We believe this means that the differences in our reductions as seen in the independent light curve fitting case are caused by differences in the choices made during the light curve fitting process, rather than in the creation of the underlying spectroscopic light curves. There are many choices, such as how to fit limb darkening, what to use for systematics modeling, whether to fit the planets together or separately, and how to deal with flares, that may affect the resulting transmission spectrum. As we have identified the likely source of our spectroscopic differences, we can look more closely into the effect each of these choices has on the results going forward. Once again, all corrected TRAPPIST-1~e spectra prefer to be fit by a flat line over the GP model with stronger evidence than in the original TRAPPIST-1~e spectra. In addition, we are able to remove the several hundred ppm spectroscopic feature that is likely stellar in origin, in all reduction's consistent light curve fitting results.

%However, the results for this test are inconclusive -- depending on the reduction used, the ratio is closer, the same, or further from a flat line in comparison with the original TRAPPIST-1~e spectrum. This is essentially equivalent to the problem with atmospheric retrievals picking up details in the spectra that may seem insignificant by eye -- even though the reductions look by eye consistent across the board, the overall specifics of each reduction have their own specific qualities across the wavelength range as a whole. Once again, the largest difference in reductions is how the flares are dealt with. Most obviously, reductions \textit{NE} and \textit{AB} only consider the integrations before the flare right after the end of the TRAPPIST-1~e transit (clearly visible in H$\alpha$ in \autoref{fig:flares}), reduction \textit{NA} fits it with a GP, while reductions \textit{CC} and \textit{KS} do not explicitly fit for the flare. These small differences in reductions, not even obvious by eye, have important implications for the further characterization of the potential atmospheric signal of TRAPPIST-1~e, especially when compounded by the need to stack many visits together. Once again, our analysis is made more difficult most obviously because of the presence of flares, especially since we do not have good models with which to directly remove the flares themselves. 

\subsection{Flares}
In \autoref{fig:wlc}, middle panel, we show the light curve for the column containing H$\alpha$ for each of our observations. H$\alpha$ is orders of magnitude more sensitive to flares than the continuum (which can be seen by the fact that the transits themselves are not even visible in NIRSpec/PRISM's H$\alpha$ bin, since it lies in a low signal region), and therefore allows us to look more carefully for flaring activity. Flares of varying magnitudes seem to be present and visible in all three observations, showing the extent of the problem flares pose for our analysis of TRAPPIST-1~e.

\paragraph{Observation 1} The large flare just before the end of the first transit is the most apparent, but smaller flares seem to be present around it as well. Most notable is a small flare during ingress of the first transit, but at least two more potential flares are seen in the later part of the observation as well. The large flare also seems to have a more complex shape to it than is visible in the white light curve, with an interruption to its expected exponential decay approximately concurrent with the transit of planet d. It could be a complex flare, or perhaps even a flare-crossing event \citep[see, e.g.,][]{Armitage_2025}.

\paragraph{Observation 3} There is a small flare right before the transit of planet e. In addition, the H$\alpha$ seems elevated at the beginning of the observation, so there may have been a larger flare that began before the beginning of the observation. This may be connected to the 4.6 $\mu$m feature seen in the spectrum ratio of this observation in \autoref{fig:ratio} (see discussion in \autoref{sec:obs3_disc}).

\paragraph{Observation 15} There are a number of relatively small flare events spread throughout the observation, including one during the transit of planet e. And again, the H$\alpha$ is high at the beginning of the observation, such that we cannot be sure how large the event preceding the visible part was, though there does not seem to be as obvious an exponential decay signal as is seen in observation 3. We count at least 5 flare peaks by eye. \\

In principle, flares break the inherent premise of our close transit technique to correct for stellar contamination, as they can change the surface of the star on timescales shorter than the time between transit observations \citep[see also][]{Vasilyev_2025}. There are large gaps in our knowledge of the characteristic behavior of flares on ultracool dwarfs in the infrared due to lack of high signal-to-noise observations pre-JWST, such that it is difficult to confidently remove them with current state of the art methods and models based on the underlying physics, and their sharp profiles and highly wavelength-dependent nature can make them difficult to detrend using methods like GPs, with resulting spectra highly dependent on the choice of fitting parameters. 

Flares represent a significant potential roadblock in attempting to detect atmospheric features on the TRAPPIST-1 planets, even if the stellar contamination can be transferred successfully between TRAPPIST-1~e and b using our program's observational setup. More complete models and simulations, of the flares themselves and potential secondary effects such as post-flare surface heating, and the corresponding spectroscopic behavior of the star, will be necessary for more in-depth characterization. However, through all of the spectroscopic observations of the TRAPPIST-1 system, including the ones presented here, we do have a large catalog of flares with which we can test and improve existing models.

\section{Conclusions}\label{sec:conclusion}
TRAPPIST-1~e presents an incredible opportunity -- to characterize a temperate terrestrial planet, but orbiting a radically different star, accessible with our current capabilities. In this paper, we present the justification and first observations from our multi-cycle JWST program that is attempting to detect an atmosphere around TRAPPIST-1~e with transmission spectroscopy, and if it has one, to determine if it has an Earth-like mean molecular weight containing CO$_2$. 

As early JWST observations of the TRAPPIST-1 system \citep{lim_2023, Radica_2024, Piaulet-Ghorayeb_2025}, including those of planet e \citep{Espinoza:2025}, show extreme stellar contamination from active surface regions that interferes with the detection of any small atmospheric signals and cannot be modeled well with current techniques, all of our transits of planet e include a close transit of the airless planet b to act as a model-independent stellar contamination proxy. We show that we would be able to detect an Earth-like atmosphere with strong significance through our proposed and accepted 15 close transit observations. This is under the assumption that the stellar contamination is able to be successfully transferred from planet b to planet e, though we show that the exact significance of atmospheric detection depends on the specifics of each noise instance as we need to stack many observations together. Our ability to detect an atmosphere hinges strongly on the presence of the 4.3 $\mu$m CO$_2$ feature, predicted as a common outcome of secondary atmosphere formation, though we are also sometimes able to significantly detect CH$_4$ as well. We obtained the first three observations from this program in July and December of 2024, with observations exactly in line with timings predicted using the modeling in \citet{Agol_2021}. However, these first observations immediately illustrate that TRAPPIST-1 presents difficulties beyond the idealized simulations. 

\subsection{The Challenge of Stellar Flares}
The most evident and problematic additional complication in our observations is the presence of flares, visible in every observation in H$\alpha$ (see \autoref{fig:wlc}, middle panel), to differing strengths and frequencies. These flares break the inherent assumption of our close transit technique that the stellar surface remains the same between the transits of planet e and b. 

While these early observations were among the longest in the program and therefore future shorter observations should statistically have fewer flares, that does not matter if a large flare happens mid-transit as in our observation 1. The flare itself inhibits our ability to accurately model the transit, seen by the much larger error bars in the spectrum of TRAPPIST-1~e in this observation, but it also might cause the stellar surface itself to change, as shown by the difference in optical stellar contamination changing from cold spot dominated in the TRAPPIST-1~e transit pre-flare to hot spot dominated in the TRAPPIST-1~b transit post-flare \citep[see also][]{Vasilyev_2025}. 

In observation 3, there also seems to have been a larger flare that occurred before the JWST observation began, evidenced by the elevated H$\alpha$ flux at the beginning of the observation. While we cannot be sure that this flare is the cause, the spectrum ratio of observation 3 shows a deviation from the expected flat spectrum ratio for a single visit around 4.7 $\mu$m present in all reductions (detected to varying significance relative to the null flat line case). The variations at these wavelengths that may be tied to the flaring activity are especially concerning due to the close proximity to the 4.3 $\mu$m CO$_2$ feature that our atmospheric detection relies on. Statistically, though, these large flares should be rare \citep[see e.g.,][]{Howard_2023}, and the more common small flares are more straightforward to fit and remove.

\subsection{Effectiveness of the Spectrum Ratio Method}
In observation 15, which does have a number of small flares but lacks the stronger decay signals seen in H$\alpha$ in observations 1 and 3, we are able to test the idea of stellar contamination transfer between the transits of planets e and b most cleanly. We find that the spectrum ratio is able to remove a component of the stellar contamination, evidenced by the ``flatter" corrected spectrum than the original spectrum, but that the final result is dependent on the specifics of our data reduction methods. This sensitivity to data reduction choices is amplified by our spectroscopic ratio technique, since it uses two transmission spectroscopic fits, with all of their underlying assumptions, to obtain one final ``corrected" spectrum. This will be important to consider as we continue in the program, obtaining and stacking many more observations together. 

As shown by the result of our light curve fitting test, the difference in the transmission spectra obtained through different reductions seems to arise from choices made during the light curve fitting process, rather than the creation of the light curves themselves. Therefore, we know we must further interrogate our light curve fitting choices to be confident in our final result, but that this difference does not come from a disagreement in the underlying data itself. However, regardless of choice of reduction or light curve fitting method, all TRAPPIST-1~e spectra corrected by taking the spectroscopic ratio with TRAPPIST-1~b are more closely described by a flat line than the original TRAPPIST-1~e spectrum. In addition, we seem to be able to remove a spectroscopic figure of several hundred ppm (though the exact feature size varies with reduction), which we attribute to stellar contamination, with the spectroscopic ratio. This leads us to believe that the use of close transits for stellar contamination correction is working, at least partially.

\subsection{Future Directions}
While we do not yet have the signal-to-noise ratio to detect atmospheric features from TRAPPIST-1~e in these observations (see \autoref{app:comp} for a quick comparison with the results from \citet{Espinoza:2025} and \citet{Glidden_2025}), we have many more observations in the works. Even if the ratio method cannot completely remove the stellar contamination from TRAPPIST-1~e, there are other ways we can use these observations to learn about the nature of this exoplanet. One example is the GP methodology for atmospheric retrievals and stellar contamination presented in \citet{Espinoza:2025}, which was able to marginalize over visit-to-visit stellar contamination signals, but not over possible persistent stellar features that might mimic exoplanet atmospheric signals (e.g., a granulation-type signature, a persistent polar spot, etc.). As discussed in that work, this possible persistent signature can be extracted from the data using the transmission spectrum of TRAPPIST-1~b: a persistent feature should be present in both, the spectrum of b and e. This sets an exciting prospect for study and analysis that we leave for future work. 

In addition, with our observations we have much more information than is currently being utilized in the analysis. We have the high-resolution stellar spectrum and out of transit slope, which is complicated by the NIRSpec instrumental slope (see \autoref{app:sys}) but in principle contains information on the active region magnitude and surface distribution during our observations. This can be combined with work on more up-to-date stellar active region models \citep[e.g.][]{Smitha_2025} to put additional constraints on stellar contamination. Flares can be tackled similarly, for which models can be self-calibrated with our observations and fit for simultaneously, more accurately removing their effects in the light curve and propagating the resulting uncertainties into our transit depth determinations, which we leave for a future work. This is a rich dataset that will not only inform us on the state of an atmosphere on TRAPPIST-1~e, but will also help us clarify the nature of active host stars. \\
\newpage
\begin{acknowledgments}
We thank our reviewer for their thoughtful comments and suggestions. This work makes use of observations made with the NASA/ESA/CSA James Webb Space Telescope. The data were obtained from MAST at STScI, which is operated by the Association of Universities for Research in Astronomy, Inc., under NASA contract NAS 5-03127 for JWST. These observations are associated with program GO 6456. Support for program GO 6456 was provided by NASA through a grant from the Space Telescope Science Institute, which is operated by the Association of Universities for Research in Astronomy, Inc., under NASA contract NAS 5-03127. The specific observations analyzed can be accessed via\dataset[DOI:10.17909/6p0w-fa37]{https://doi.org/10.17909/6p0w-fa37}. NHA acknowledges support by the National Science Foundation Graduate Research Fellowship under Grant No. DGE1746891. CIC acknowledges support by NASA Headquarters through an appointment to the NASA Postdoctoral Program at the Goddard Space Flight Center, administered by ORAU through a contract with NASA. DRL acknowledges support from NASA under award number 80GSFC24M0006. This material is based upon work supported by NASA under Agreement No.\ 80NSSC21K0593 for the program ``Alien Earths''.
The results reported herein benefited from collaborations and/or information exchange within NASA’s Nexus for Exoplanet System Science (NExSS) research coordination network sponsored by NASA’s Science Mission Directorate. GTM acknowledges support by the National Science Foundation MPS-Ascend Postdoctoral Research Fellowship under Grant No. 2402296.

\end{acknowledgments}

%% To help institutions obtain information on the effectiveness of their 
%% telescopes the AAS Journals has created a group of keywords for telescope 
%% facilities.
%
%% Following the acknowledgments section, use the following syntax and the
%% \facility{} or \facilities{} macros to list the keywords of facilities used 
%% in the research for the paper.  Each keyword is check against the master 
%% list during copy editing.  Individual instruments can be provided in 
%% parentheses, after the keyword, but they are not verified.

\vspace{3mm}
\facilities{JWST(NIRSpec)}

%% Similar to \facility{}, there is the optional \software command to allow 
%% authors a place to specify which programs were used during the creation of 
%% the manuscript. Authors should list each code and include either a
%% citation or url to the code inside ()s when available.

\software{astropy \citep{2013A&A...558A..33A,2018AJ....156..123A, AstropyCollaboration_2022}, batman \citep{batman}, corner \citep{corner}, celerite \citep{celerite}, dynesty \citep{dynesty}, emcee \citep{emcee}, Eureka \citep{eureka}, ExoTiC-JEDI \citep{exoticjedi}, ExoTiC-LD \citep{Grant2024}, george \citep{hodlr}, juliet \citep{juliet}, jupyter \citep{Kluyver2016jupyter}, matplotlib \citep{Hunter:2007}, multinest \citep{Feroz_2008, Feroz_2009, Feroz_2019}, NumPy \citep{harris2020array}, SciPy \citep{2020SciPy-NMeth}, seaborn \citep{seaborn}, transitspectroscopy \citep{transitspectroscopy}}

%% Appendix material should be preceded with a single \appendix command.
%% There should be a \section command for each appendix. Mark appendix
%% subsections with the same markup you use in the main body of the paper.

%% Each Appendix (indicated with \section) will be lettered A, B, C, etc.
%% The equation counter will reset when it encounters the \appendix
%% command and will number appendix equations (A1), (A2), etc. The
%% Figure and Table counter will not reset.

\clearpage
\bibliography{bib}{}
\bibliographystyle{aasjournal}

%% This command is needed to show the entire author+affiliation list when
%% the collaboration and author truncation commands are used.  It has to
%% go at the end of the manuscript.
%\allauthors

%% Include this line if you are using the \added, \replaced, \deleted
%% commands to see a summary list of all changes at the end of the article.
%\listofchanges

\appendix

\section{Simple test: Can we detect an Earth-like CO$_2$ feature?}\label{sec:simple}
\begin{figure}[h]
    \centering
    \includegraphics[width=0.5\linewidth]{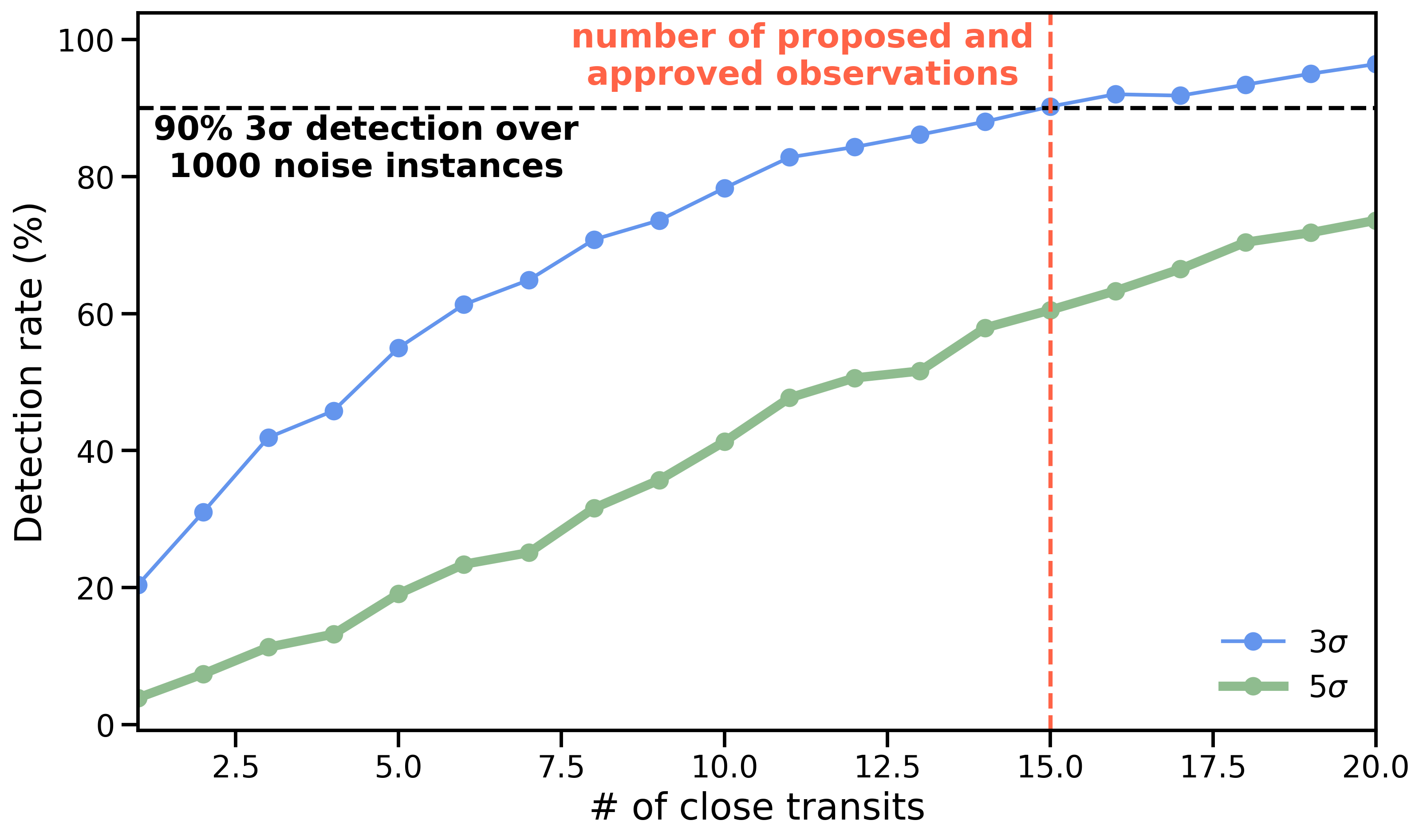}
    \caption{Detection rate of the 4.3 $\mu$m CO$_2$ feature over 1000 noise realizations as a function of number of close transits of TRAPPIST-1~e and b, considering from 1 to 20 transits. We show curves corresponding to a $3\sigma$ and a $5\sigma$ detection of an atmosphere, which we are able to detect 90\% and $\sim$60\% of the time over the 1000 cases, respectively.}
    \label{fig:simple-detect}
\end{figure}

Our understanding of the formation and evolution of rocky planet atmospheres predicts CO$_2$ as a common constituent species \citep[see e.g.][]{Herbort_2020}, which is also the strongest predicted spectroscopic feature present in the NIRSpec/PRISM bandpass. In addition, CO$_2$ is not present in stellar surface features, while other potential molecules like H$_2$O (directly present in M dwarf spots) or CH$_4$ \citep[similar opacity seen created by stellar contamination in some past observations,][]{Espinoza:2025} are somewhat more dangerous to rely on from a stellar contamination perspective \citep[see e.g., models in][]{Rackham_2018}. Therefore, for our most conservative scenario, we estimate our ability to detect the 4.3 $\mu$m CO$_2$ feature by measuring the average transit depth in (4.2-4.4 $\mu$m) and out (3.7-4.1 $\mu$m) of the feature. Using the same atmospheric and stellar modeling and resulting simulated transmission spectra using the same spectrum binning and empirical errors as in \citet{Espinoza:2025} as used for \autoref{sec:complex}, we consider from 1 to 20 close transits, averaging the spectrum and resulting errors as we add more transits together, and repeat the experiment 1000 times. \autoref{fig:simple-detect} shows our detection percentage of those 1000 trials as a function of number of close transits for a $3\sigma$ and a $5\sigma$ detection. With 15 close transit observations, we are able to detect an Earth-like CO$_2$ feature 90\% of the time at $3\sigma$ (where $\sigma$ is determined through the out of feature spectrum depth variance), and approximately 60\% of the time at $5\sigma$. 

We also carry out a test for a false positive atmospheric detection using this method (i.e. detecting an atmosphere when none is present) by repeating the same experiment as above, but with flat lines representing both TRAPPIST-1~e and b's spectra. For 15 close transits, this is well below 10\% at the $3\sigma$ level. 

\section{Full Simulated Retrieval Posterior}\label{app:corner}
\begin{figure*}[h!]
    \centering
    \includegraphics[width=0.8\linewidth]{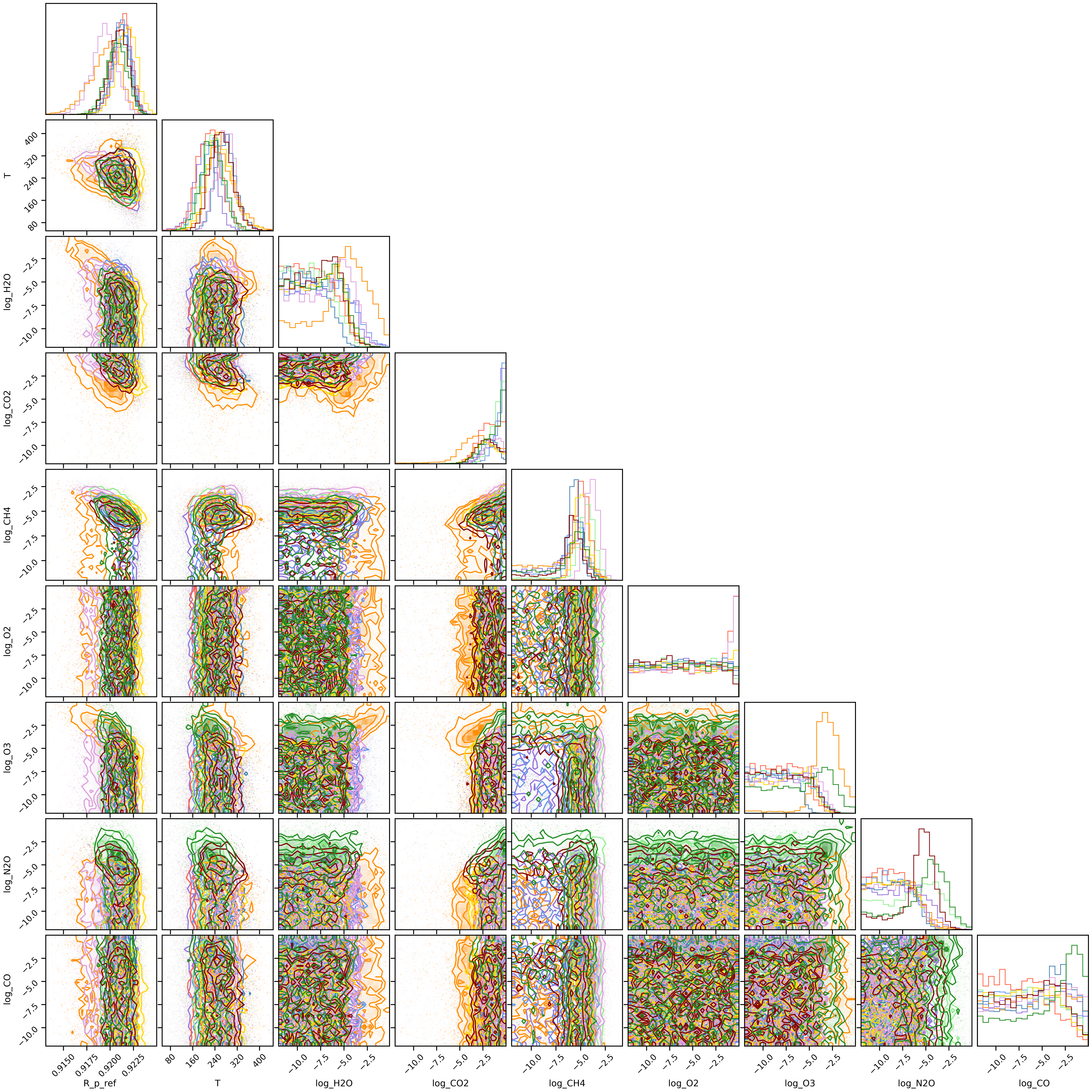}
    \caption{The posterior distributions associated with the 15 close transit retrievals for each of the 10 noise instances shown in \autoref{fig:complex-detect}. %The narrow posteriors in black correspond to zero noise retrievals, essentially the limit of detection significance given observations at the resolution of NIRSpec/PRISM and therefore taken as the ``ground truth" values. 
    We confidently detect CO$_2$ in all noise instances, and sometimes are able to detect CH$_4$ as well. None of the other species are constrained to greater than upper limits.}
    \label{fig:corner_full}
\end{figure*}

We show in \autoref{fig:corner_full} the full posterior with all retrieved species for the simulated retrieval result shown in \autoref{fig:complex-detect}.

\section{Absolute transit depths}\label{app:absolute}
\begin{figure*}[h!]
    \centering
    \includegraphics[width=0.9\linewidth]{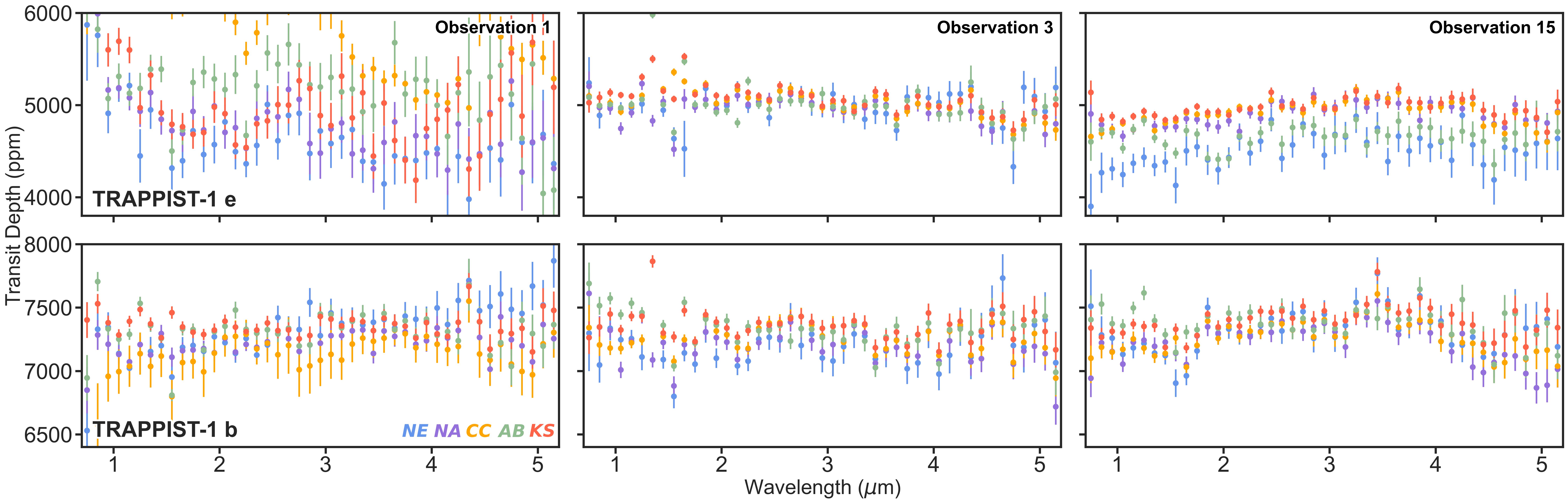}
    \caption{Same as \autoref{fig:wlc}, bottom panel, but in absolute transit depth. Most of the reductions show very similar absolute transit depths, but both observation 1 spectra and the TRAPPIST-1~e spectrum from observation 15 show clear offsets due to the handling of flares.}
    \label{fig:spec_abs}
\end{figure*}

\begin{figure*}
    \centering
    \includegraphics[width=0.9\linewidth]{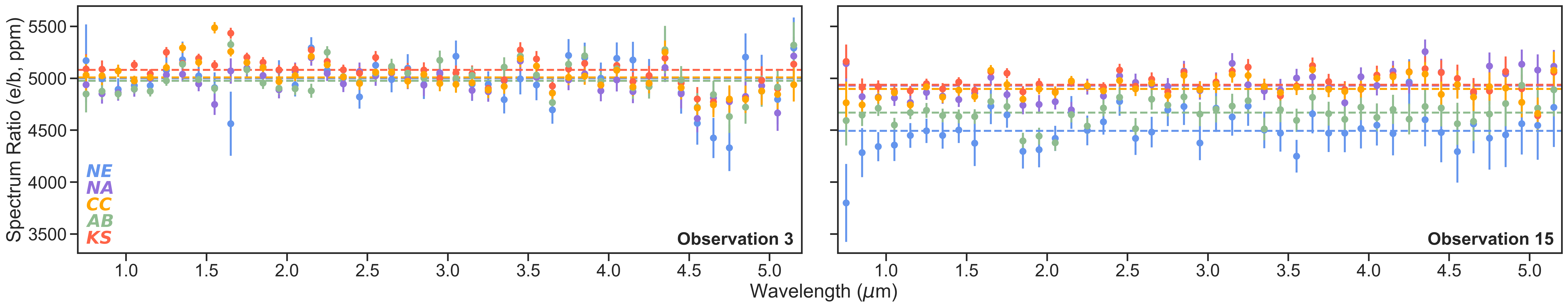}
    \caption{Same as \autoref{fig:ratio}, but in absolute spectrum ratio. The dashed lines of each color represent the median spectrum ratio for each reduction. Observation 15 shows an offset in absolute spectrum ratio depth due to the difference of treatment of flares, which can also be seen in the individual TRAPPIST-1~e spectrum in \autoref{fig:spec_abs}.}
    \label{fig:ratio_abs}
\end{figure*}

We show in \autoref{fig:spec_abs} the same spectra shown in \autoref{fig:wlc}, bottom panel, but in absolute transit depth rather than median subtracted, to enable a direct transit depth comparison between reductions and visits. We also show in \autoref{fig:ratio_abs} the same spectrum ratios shown in \autoref{fig:ratio}, but in absolute spectrum ratio. As a reminder, the wavelength region from approximately 1.1-1.7 $\mu$m is saturated, which is the cause of the extra scatter seen in this range in some observations. Our analyses of planet e's spectrum have two different solutions families, with a vertical offset between them. This seems to be caused by the choice of data used in the fit: transit fits from \textit{NE} and \textit{AB} only use the data up to the flare shortly after the planet's egress, while the rest of the analyses instead choose to detrend the flare and include more post-transit baseline (see \autoref{sec:red} for details). This interpretation is supported by testing the \textit{NA} reduction using the same individual transit integration cuts as used in the \textit{NE} reduction, for which a consistent spectrum was obtained between the \textit{NA} and \textit{NE} reductions. All spectra agree when median-subtracted though, which means the spectral shape/features themselves are unchanged by the choice of included baseline. It is interesting to note that the absolute spectrum ratio values differ slightly between the two visits.

\section{TRAPPIST-1~d transit}\label{app:d}
\begin{figure*}[h!]
    \centering
    \includegraphics[width=0.9\linewidth]{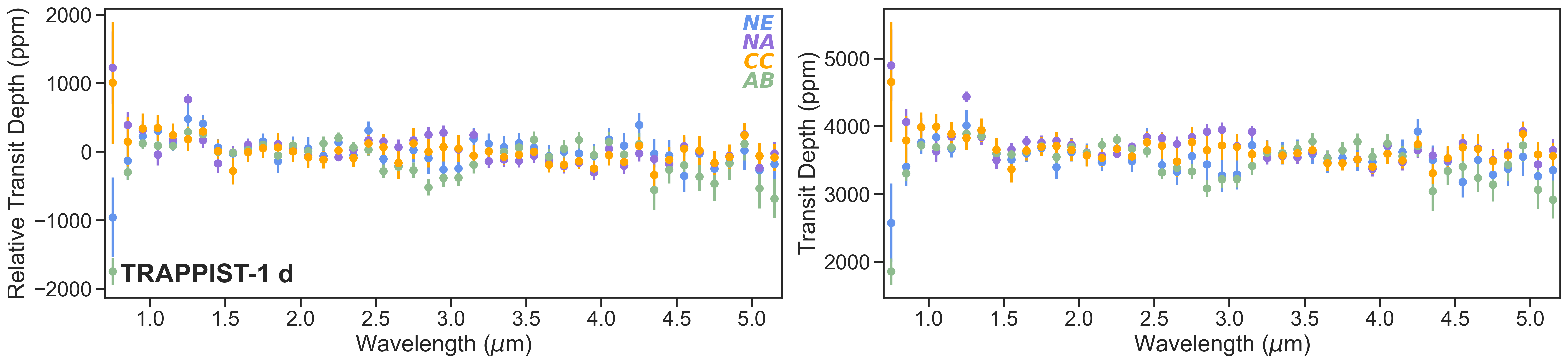}
    \caption{Spectrum of TRAPPIST-1~d from observation 1, in both relative transit depth (left) and absolute transit depth (right). The slope in the optical is different between reductions, due to the method of handling the large flare.}
    \label{fig:d}
\end{figure*}

There is an extra transit of TRAPPIST-1~d that appears in observation 1. As it is concurrent with the large flare in this observation, the resulting transmission spectrum is highly dependent on the method used to remove the out of transit trend, especially in the shortest wavelengths. We show a comparison of reductions in \autoref{fig:d}, both in relative and absolute transit depth, though note there is no \textit{KS} reduction.
\paragraph{NE} Transit was reduced and fit the same way as the rest of the observation 1 transits detailed in \autoref{sec:red_ne}, using integrations 4715 to 9925.
\paragraph{NA} Transit was reduced and fit the same way as the rest of the observation 1 transits detailed in \autoref{sec:red_na}, using integrations 5000 to 10000.
\paragraph{CC} Transit was reduced and fit concurrently with the other observation 1 transits detailed in \autoref{sec:red_cc}.
\paragraph{AB} Transit was reduced and fit the same way as the rest of the observation 1 transits detailed in \autoref{sec:red_ab}, using integrations 4800-10000.

\section{NIRSpec/PRISM instrumental slope}\label{app:sys}
There is a well-known systematic slope in the NRS1 detector, which NIRSpec/PRISM illuminates. Typically this slope is treated as a nuisance and is simply removed as a function of wavelength in the light curve fitting process. However, we can use the fit slopes as a function of wavelength to get a ``spectrum" of the slope itself. 

\begin{figure*}[h!]
    \centering
    \includegraphics[width=0.9\linewidth]{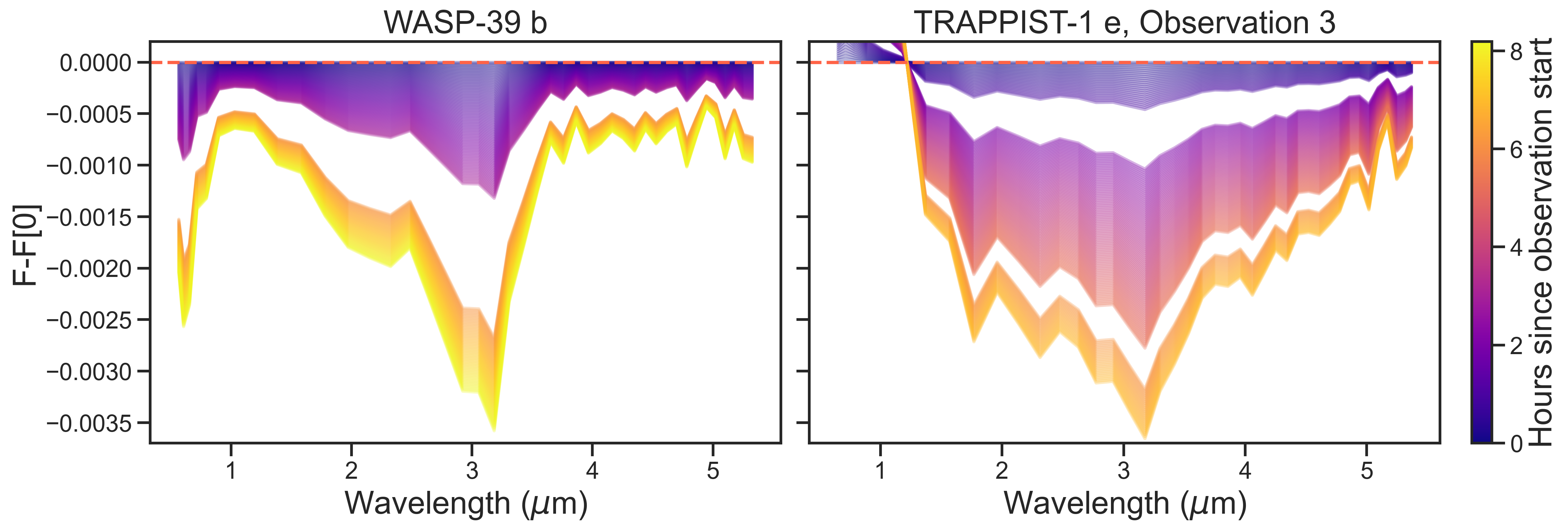}
    \caption{Change in out of transit linear slope as a function of time, shown as the change in relative flux in reference to the beginning of the observation, for WASP-39 (left) and TRAPPIST-1, observation 3 (right). Each curve is colored by the time since the beginning of the observation, on the same scale. While parts of the change in out of transit slope look similar, therefore likely pointing to an instrumental origin, TRAPPIST-1~e has a stronger change in out of transit linear slope across the shown wavelength range, likely due to the addition of a stellar component to the out of transit slope. Breaks in the curves are due to the masking of the transit events.}
    \label{fig:instrumental}
\end{figure*}

Using the \textit{transitspectroscopy} \textit{NA} reduction, we take the linear component of the systematics model made during the spectroscopic light curve fits, which contains the bulk slope present in the light curves. We average over 100 time integrations and over $10\times$ pixels to increase the signal in each calculated slope curve. The change in this out of transit linear slope as a function of time (meaning the difference between each averaged slope curve with the first slope curve) is shown in \autoref{fig:instrumental}. However, TRAPPIST-1 presents a complicated scenario, as this out of transit slope is likely not only systematic in nature, but also contains a component due to rotational modulation of the active host star. While there are clues to this in our own observations, we also have independent evidence from NIRISS/SOSS observations of the TRAPPIST-1 planets, since NIRISS/SOSS has a much smaller instrumental slope \citep[e.g.,][]{Radica_2023}. Observations of TRAPPIST-1~b from \citet{lim_2023} show significant out of transit light curve modulations, on the same timescale as our observations presented here. 

Therefore, we also perform this same analysis on the WASP-39~b NIRSpec/PRISM observations \citep{Rustamkulov_2023}. WASP-39 is known as an inactive star, and the observations from multiple wavelengths and instruments from the JWST ERS program \citep{Ahrer_2023, Alderson_2023, Feinstein_2023} do not show any evidence of rotational modulation, which suggests that we should not see a stellar component to the out of transit slope. The length of the WASP-39~b NIRSpec/PRISM observation is also approximately the same as the length of the TRAPPIST-1 observation 3 presented here (8.4 vs 7.1, respectively), which allows for a direct comparison in the strength of the instrumental slope over time. We reduce this data in the same way as described in \autoref{sec:red_na}, and show the same linear systematics component in \autoref{fig:instrumental}.

The overall shape of each out of transit slope plot is similar, but clearly differs. This is also similar to the shape of the NRS1 slope from G395H observations of HAT-P-14~b from JWST commissioning \citep{Espinoza_2023}. The strength of the two slopes and their shape from approximately 1 to 3 $\mu$m is comparable, but the TRAPPIST-1~e slope is stronger, and over a slightly shorter time (1.3 hour shorter observation). In addition, the out of transit slope is much stronger longwards of 3 $\mu$m, and also is in the opposite direction shortwards of 1 $\mu$m. This extra component not seen in WASP-39~b is likely due to the added contribution of a slightly changing stellar signal throughout the observation, either due to flares or rotational modulation. Disentangling the instrumental and astrophysical components of this signal will be difficult, as it is unknown what the instrumental slope is sensitive to (i.e. if it changes in time or as a function of incident stellar SED), and as such is beyond the scope of this work. A detailed investigation of the constraints on astrophysical variations that are possible in light of the systematics will be the focus of a future study. %There is no consistent noise seen in these slopes from 4.4-4.9 $\mu$m, which is where the gaussian feature appears in \autoref{fig:gauss}, so it is unlikely that the feature is instrumental in origin. However, there is a feature at slightly longer wavelengths in both observations, though this is not causing 

\section{Stellar Contamination Retrievals}\label{app:ret_fits}
\begin{figure*}[h!]
    \centering
    \includegraphics[width=\linewidth]{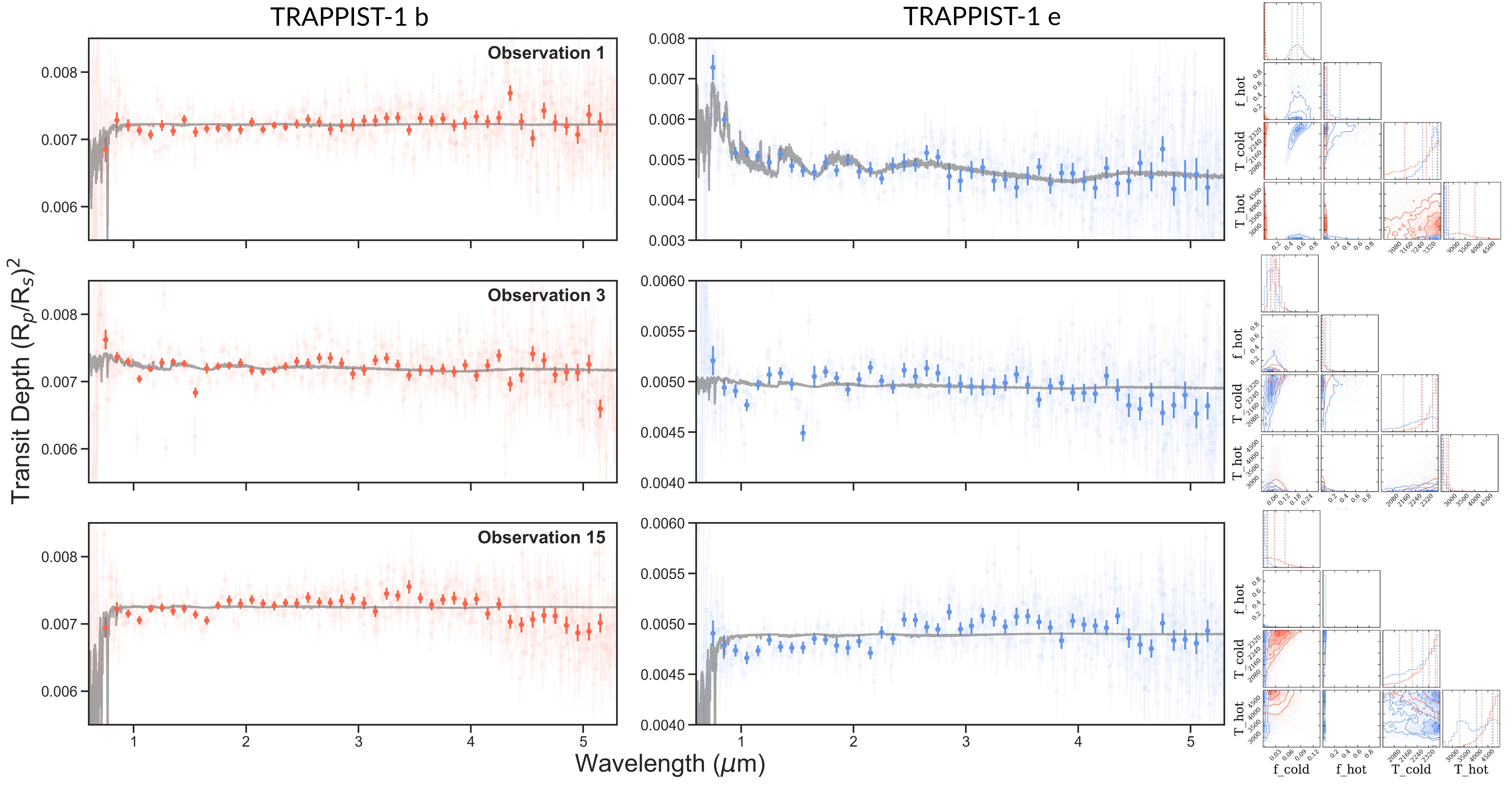}
    \caption{Stellar contamination retrieval models for all of our TRAPPIST-1~b and e observations, with the stellar contamination parameter posteriors on the right. For observation 1, the two transits show clearly different stellar contamination signatures, which is also reflected in the posteriors, with TRAPPIST-1~b dominated by a hot component post-flare and TRAPPIST-1~e dominated by a cold component pre-flare. The other visits show common retrieved stellar contamination parameters, but overall the models provide poor fits to the observed spectra.}
    \vspace{-5mm}
    \label{fig:sc}
\end{figure*}

We carried out stellar contamination retrievals with POSEIDON, by assuming that some fraction of the photosphere is covered in a photospheric model of another temperature, representative of the spot temperature. We assume no atmospheric signal, as any potential atmospheric signal would be beneath the noise from any individual visit, and therefore model the entire spectrum as due to stellar contamination and only fit for a reference radius for the planet, representative of the average transit depth. We use the \textit{NA} reduction for our tests, and perform the retrievals on the pixel-level spectrum. For all retrievals, we fit for photospheric temperature with a normal distribution [2566, 26] (mean, $\sigma$) from \citet{Agol_2021}, and fit for cold spot temperatures uniformly from [2000, 2400] (PHOENIX models, used for stellar contamination retrievals in POSEIDON, do not go colder than 2000 K) and hot spot temperatures uniformly from [2600, 5000]. Cold and hot spot covering fractions are allowed from 0 (none) to 1 (full). We do not perform retrievals on the transmission spectrum of TRAPPIST-1~d as the optical slope, the most clear evidence of stellar contamination signal, varies between reductions due to different techniques for detrending the flare (see \autoref{fig:d}). All retreival results are shown in \autoref{fig:sc}.

In observation 1, it is clear that the TRAPPIST-1~b and e spectra are affected by different stellar contamination signatures. The b spectrum, which is after the large flare in this observation, shows evidence of a hot spot dominated surface, while the pre-flare TRAPPIST-1~e instead shows strong contamination from cold spots. On the other hand, observations 3 and 15 show consistent retrieved parameters between their b and e transmission spectra. However, overall the models are a poor fit to the observations -- there are clearly deviations from a flat line caused by stellar activity that are not captured by the models. This is not unexpected, and is the motivation for the model-independent method of stellar contamination removal that our JWST program relies on. 

\section{Light Curve Fitting Test}\label{app:fitting_test}
\begin{figure}
    \centering
    \includegraphics[width=0.4\linewidth]{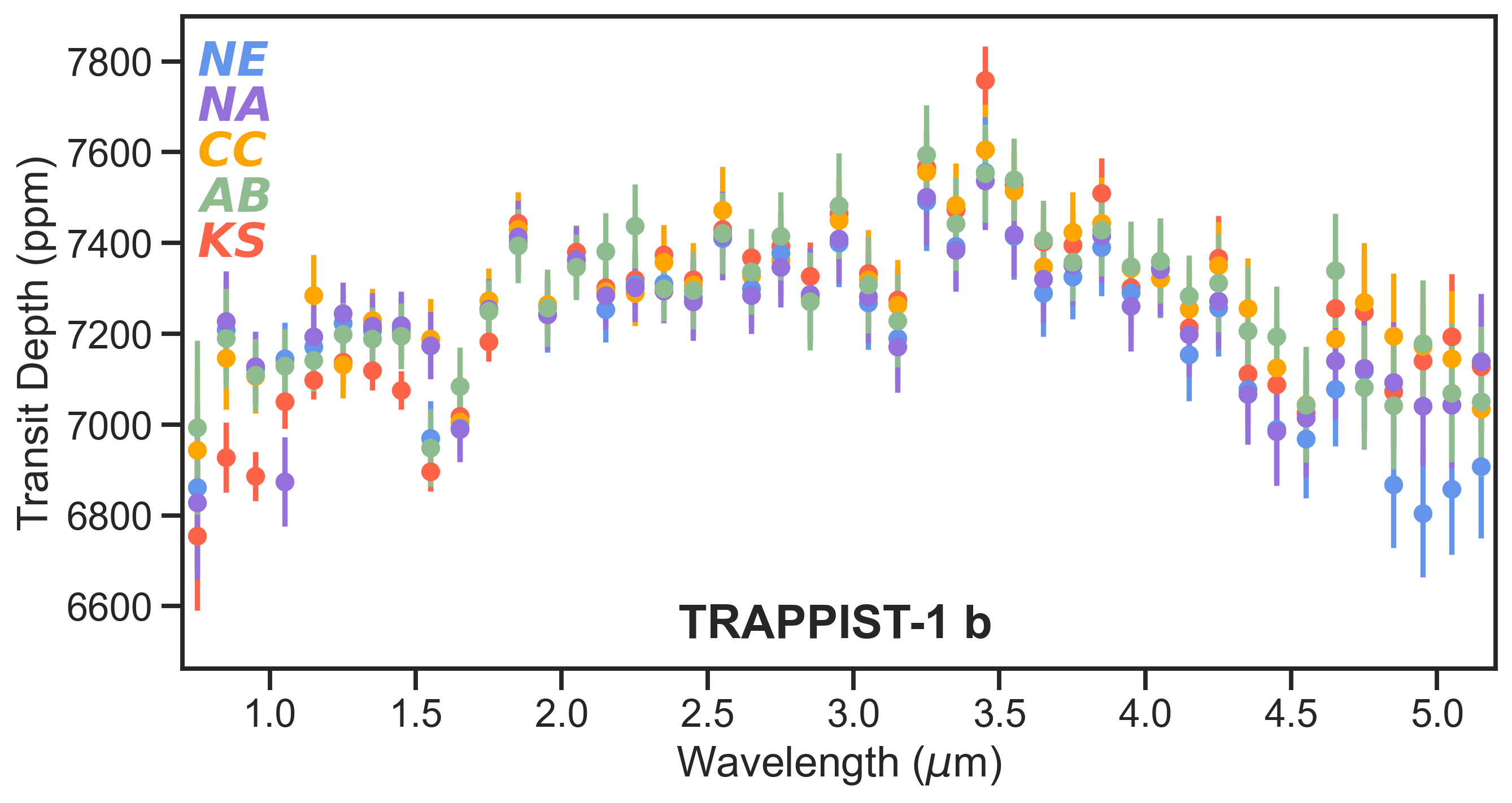}
    \includegraphics[width=0.4\linewidth]{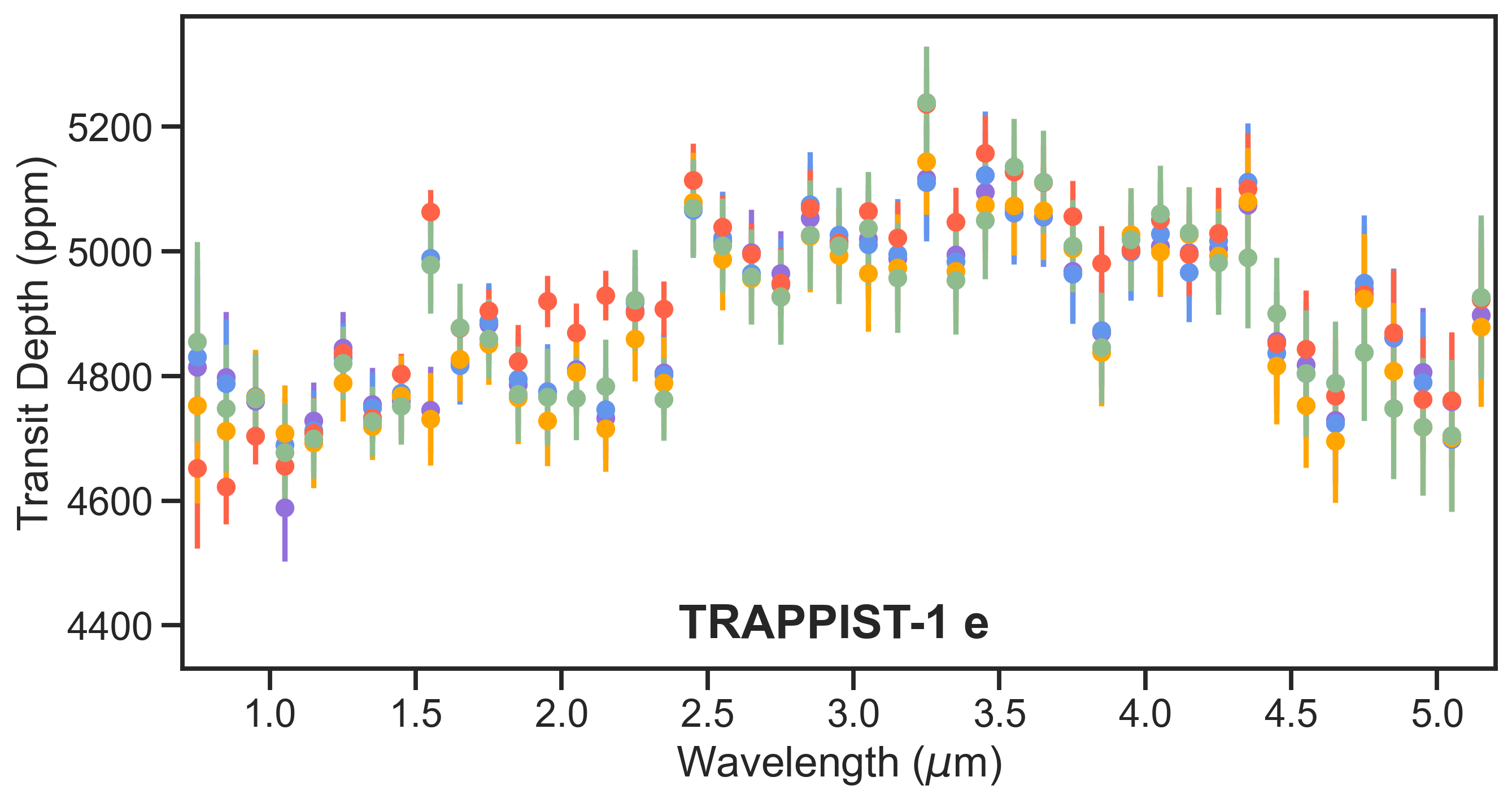}
    \includegraphics[width=0.4\linewidth]{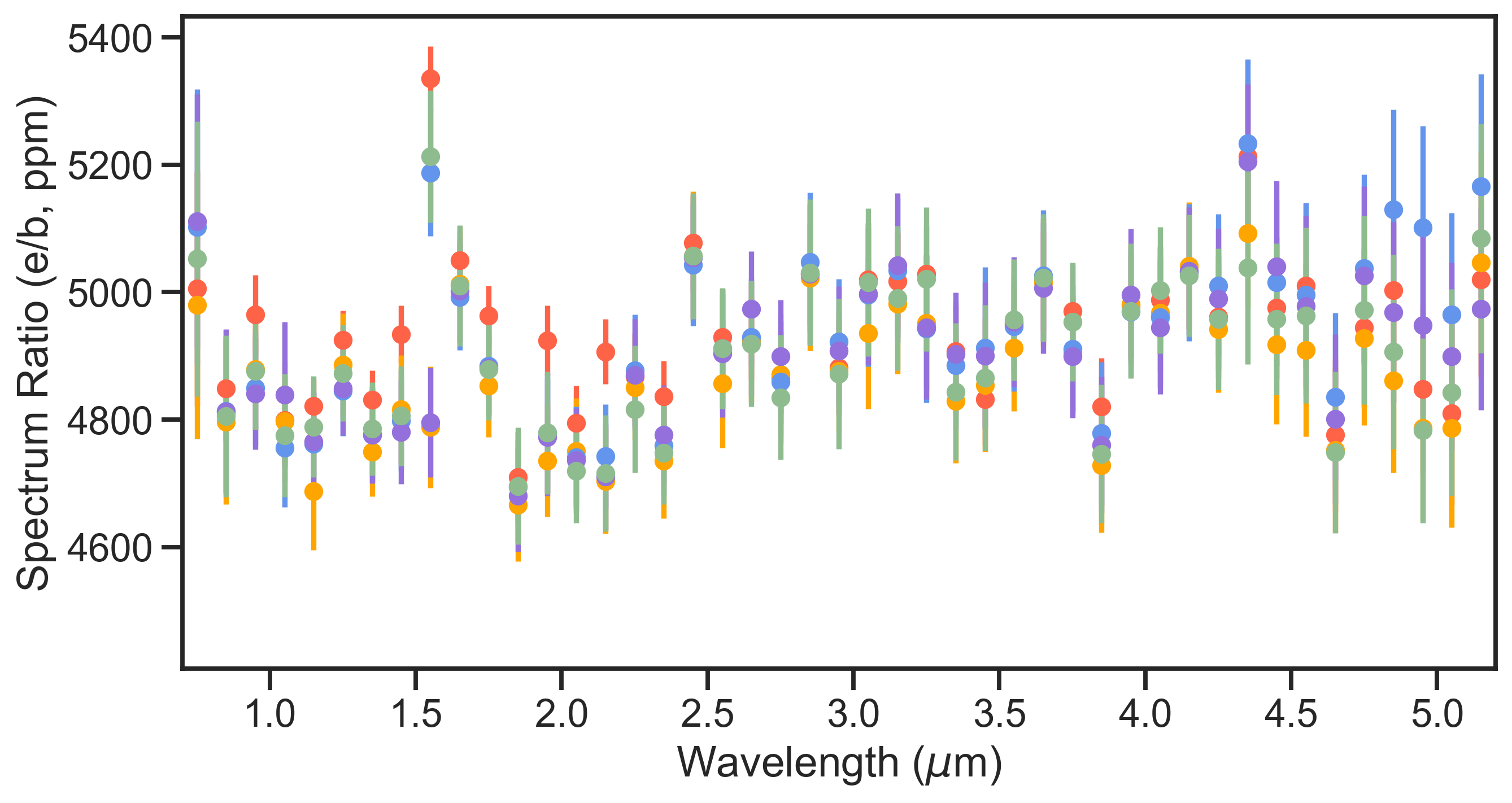}
    \caption{The TRAPPIST-1~b and TRAPPIST-1~e transmission spectra, along with the resulting spectroscopic ratio, for the consistent light curve fitting test. All reductions show close agreement, suggesting that the differences in spectra seen in \autoref{fig:wlc} is due to choices made during the light curve fitting process rather than the generation of the light curves from the raw data.}
    \label{fig:fitting_test}
\end{figure}

We show in \autoref{fig:fitting_test} the TRAPPIST-1~e, TRAPPIST-1~b, and spectrum ratio for the consistent light curve fitting test described in \autoref{sec:obs15_disc} for observation 15. It can be seen that there is remarkable agreement between the spectra from the different reductions, which suggests that the small, but still potentially significant, differences between the independently produced spectra come from differences in the light curve fitting process rather than a difference in the generation of the underlying spectroscopic light curves. 

\section{Comparison to TST-DREAMS observations}\label{app:comp}
\begin{figure*}
    \centering
    \includegraphics[width=\linewidth]{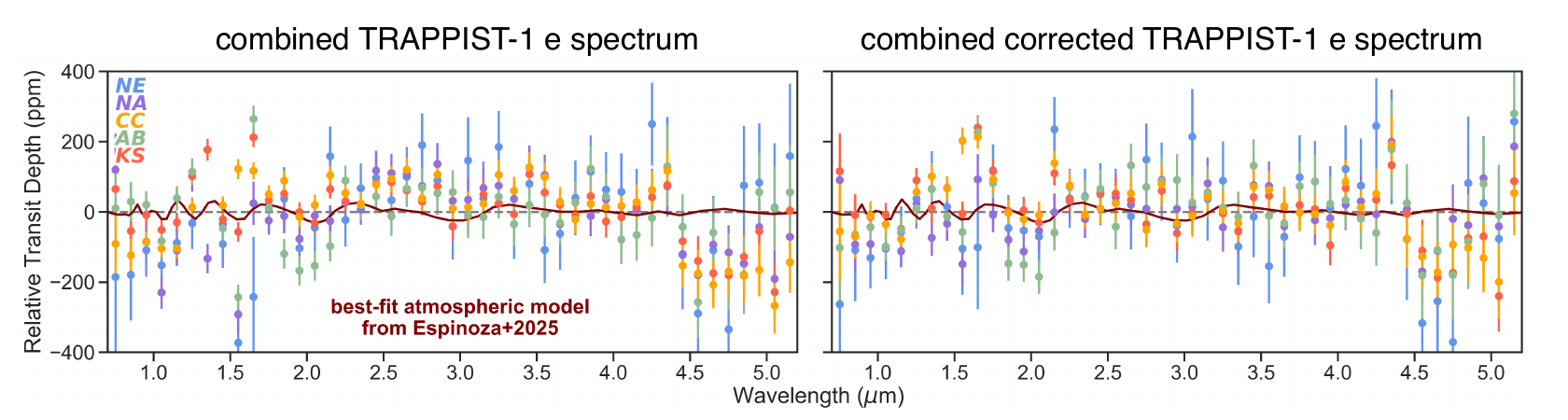}
    \caption{Combined TRAPPIST-1~e spectrum (left) and combined, corrected TRAPPIST-1~e spectrum (right), both from visits 3 and 15. Plotted for comparison is the best-fit atmospheric model from \citet{Espinoza:2025}, for reference. The model lies well within the error bars of our two clean visits.}
    \label{fig:comb}
\end{figure*}

We leave a full comparison to the results from \citet{Espinoza:2025} and \citet{Glidden_2025}, including atmospheric retrievals, until we have comparable signal to those four transit observations. However, for reference, we show in \autoref{fig:comb} the combined TRAPPIST-1~e spectrum, both corrected and not, from observations 3 and 15, along with the best-fit potential atmospheric model from \citet{Espinoza:2025}. This model was obtained by using a GP to fit for the stellar contamination simultaneously with an atmospheric retrieval with POSEIDON. A flat line and the persistent signal model from \citet{Espinoza:2025} are equally good fits to the data according to a chi square fit, which confirms that with only these three observations we cannot confirm nor reject this signal. %The model is consistent with our current observations, though with two transits, our data uncertainties are not to the point of direct comparison ().

The observations obtained through this program will also allow for a direct test of this persistent signal obtained through the four observations in \citet{Espinoza:2025} and \citet{Glidden_2025}. The persistent signal seen through those four observations could be either due to a common stellar signal that is always present on the surface (such as something analogous to surface granulation), or could be due to a planetary atmosphere signal. If we see the same features appear again in our program's observations of TRAPPIST-1~e, which must be the case if the signal may be planetary in origin, we will be able to test for the possible stellar origin through performing the exact same analysis on the airless TRAPPIST-1~b. If the signal is stellar in origin, it is likely to be seen in the TRAPPIST-1~b spectrum as well, but will not be present if the signal is truly originating from TRAPPIST-1~e.

\end{document}